\documentclass[10pt, onecolumn, notitlepage]{revtex4-2}

\usepackage{amsfonts}       
\usepackage{dsfont}			
\usepackage{braket}			
\usepackage{amsmath}		
\DeclareMathOperator{\tr}{tr}
\newtheorem{thm}{Theorem}
\newtheorem{cor}{Corollary}

\usepackage{float}			
\usepackage{tikz}
\usepackage{pgfplots}		
\usepackage{adjustbox} 		
\usetikzlibrary{calc,shapes.geometric,backgrounds,positioning,shapes.geometric,decorations.markings,decorations.pathreplacing,arrows,knots,hobby,angles,quotes}
\usetikzlibrary{arrows,shapes,snakes,automata,backgrounds,petri}

\usepackage{todonotes}		
\usepackage{subcaption}

\definecolor{color1}{HTML}{63acbe} 
\definecolor{color2}{HTML}{ee442f} 
\definecolor{color3}{HTML}{601a4a} 
\definecolor{color4}{HTML}{fef7f6} 

\begin{document}
	\title{Quantum machine learning of graph-structured data} 
	\author{Kerstin Beer}
	\email{kerstin.beer@itp.uni-hannover.de}
	\author{Megha Khosla }
	\author{Julius Köhler}
	\author{Tobias J.\ Osborne}
	\affiliation{Institut f\"ur Theoretische Physik, Leibniz Universit\"at Hannover, Germany}
	\affiliation{L3S Research Center, Leibniz Universit\"at Hannover, Germany}
	\maketitle

	\maketitle
	
	\textbf{
		Graph structures are ubiquitous throughout the natural sciences. Here we consider graph-structured quantum data and describe how to carry out its \emph{quantum machine learning} via \emph{quantum neural networks}. In particular, we consider training data in the form of pairs of input and output quantum states associated with the vertices of a graph, together with edges encoding correlations between the vertices. We explain how to systematically exploit this additional graph structure to improve  quantum learning algorithms. These algorithms are numerically simulated and exhibit excellent learning behavior. Scalable quantum implementations of the learning procedures are likely feasible on the next generation of quantum computing devices.   
	}
	\section{Introduction}
	
	With the experimental advent of large-scale quantum computation \cite{Google2019}  we are entering the \emph{quantum information era}. As we transition this epoch, \emph{noisy intermediate-scale quantum devices} (NISQ) will play a central role in processing quantum information \cite{preskillQuantumComputingNISQ2018a}, introducing challenges and opportunities throughout the natural sciences; a NISQ device is capable of processing quantum information. However noise and imperfections prevent full scalable fault-tolerant quantum computation. Leading NISQ device vendors will likely offer, in the coming year, quantum solutions routinely comprising more than 100 qubits. 
	
	The coming ubiquity of NISQ information processing devices brings with it an extraordinary situation for physics and computer science alike, heralding for the first time the routine production of large-scale correlated (yet noisy) quantum states of hundreds of qubits. The characterization of such \emph{quantum data sources} currently occupies a large amount of classical computational effort because the information required to classically learn and specify a quantum state via \emph{tomography} \cite{nielsenQuantumComputationQuantum2000} scales exponentially with the number of qubits. 
	
	Full tomography of the complex quantum states produced by extant NISQ devices is actually already entirely out of the question. To address the critical challenge of learning unknown noisy states of many qubits, we must take recourse to approximate inference methods such as \emph{compressed sensing} \cite{grossQuantumStateTomography2010,flammiaQuantumTomographyCompressed2012} or develop entirely new technologies. We believe that full \emph{quantum machine learning} (QML) \cite{Biamonte2017,cilibertocarloQuantumMachineLearning2018,schuldQuantumMachineLearning2017}, where quantum devices themselves are exploited to learn quantum data, offers the most promising solution to the crucial problem of learning and characterizing complex quantum data sources. 
	
	Quantum machine learning, whereby classical ML is generalized to the quantum realm, has enjoyed a recent renaissance, leading to a dizzying array of formulations and applications (see \cite{Biamonte2017,cilibertocarloQuantumMachineLearning2018,schuldQuantumMachineLearning2017} and references therein for a cross section). Broadly speaking one has the following taxonomy \cite{aimeurMachineLearningQuantum2006}: (i) quantum speedups for classical ML \cite{aimeurQuantumSpeedupUnsupervised2013,paparoQuantumSpeedupActive2014,schuldQuestQuantumNeural2014,kapoorQuantumPerceptronModels2016a}; (ii) classical ML to characterize quantum systems \cite{Lovett2013,carleoSolvingQuantumManybody2017,tierschAdaptiveQuantumComputation2015}; or (iii) quantum devices to learn quantum data (``full'' QML) \cite{sasakiQuantumLearningUniversal2002,gambsQuantumClassification2008,SCMB12,DTB16,monrasInductiveSupervisedQuantum2017a,Alvarez2017,Amin2018,Du2018,SMMCB19,beerTrainingDeepQuantum2020,verdonUniversalTrainingAlgorithm2018}. Our focus here is on the last category, as it is this scenario where quantum speedups are not only most likely, but also most urgently required owing to the aforementioned exponential difficulty of tomography.
	
	A variety of quantum architectures for QML have been considered, from variational quantum circuits \cite{Du2018,farhiQuantumAlgorithmsFixed2017} to quantum analogues of artificial neural networks \cite{DTB16,Alvarez2017,Amin2018,SMMCB19,DB18,beerTrainingDeepQuantum2020}. We believe that the quantum neural network (QNN) architecture introduced in \cite{beerTrainingDeepQuantum2020} offers a most promising platform for full QML. For example, such QNNs have been recently exploited as \emph{quantum autoencoders} to carry out the denoising of entangled quantum states \cite{bondarenko2020quantum}. Additionally, these QNNs appear to offer an architecture -- when the quantum neurons are sufficiently local and sparse \cite{sharmaTrainabilityDissipativePerceptronBased2020} -- which might potentially be exploited to avoid the ``barren plateaux'' problem \footnote{The so-called \emph{barren plateaux} is a manifestation of the vanishing gradient problem which appears specific to the quantum setting \cite{mccleanBarrenPlateausQuantum2018}.}. Finally, these QNNs have been found to reach the fundamental information-theoretic limits on quantum learning \cite{arunachalamGuestColumnSurvey2017,gammelmarkQuantumLearningMeasurement2009,sasakiQuantumTemplateMatching2001,sasakiQuantumLearningUniversal2002,sentisQuantumLearningQuantum2012,monrasInductiveSupervisedQuantum2017a} imposed by the \emph{quantum no free lunch theorem} \cite{polandNoFreeLunch2020, sharma2020reformulation, bisioOptimalQuantumLearning2010a},
	a bound on the performance of quantum learning of generic unstructured quantum data sources.     
	
	Quantum data sources will never be generic and unstructured because the devices producing them always have structure. Indeed, causal and spatial order manifest themselves in correlations between the states produced by nearby local data sources. So it is that physics is even possible: without causal locality, we could never have characterized the laws of physics! To quantify such correlations it is most convenient to introduce a graph structure via a finite (or infinite) graph $G = (V,E)$, where $V$ denotes the set of vertices and $E$ the set of edges.
	
	There have already been some investigations exploiting graph structure for QML \cite{cong2019quantum,verdon}. Here the emphasis has so far been on building the graph structure into the neural network ansatz itself. However, a critical open challenge facing QML is to teach a complex QNN the \emph{a priori} variable graph structure of the quantum source itself. Here an approach that bakes the graph adjacency structure into the variational network ansatz faces difficulties. It is the crucial challenge of exploiting a quantum source's graph  structure to improve QML with an \emph{arbitrary} QNN, which we take aim at here: our main contribution is a general method to improve the learning efficiency -- and the generalization behaviour -- of QML via an arbitrary QNN ansatz, by exploiting graph structure.  
	
	The archetypal problem we consider here is that of a distributed set of quantum information processors, associated with the vertices of a graph $G$. A processor at vertex/site $j$ takes as input a state $\rho_j$. The edges $E$ of the graph encode the correlations induced between, e.g., by the spatial vicinity, these processors. The goal is to optimally learn input-output relations for this distributed set of processors: we are given a training set $\{(\rho_j, \sigma_j)\,|\, j = 1,2, \ldots, S\}$ of ideal outputs $\sigma_j$ corresponding to an input $\rho_j$ for a processor at vertex/site $j$. Such a scenario flexibly models a wide variety of physically relevant situations ranging from distributed networks of atomic clocks through to quantum NISQ device clusters.  	
	
	In this paper, we initiate the study of graph-structured quantum data sources. Our emphasis is on learning and characterizing the graph structure of \emph{noisy and unreliable} quantum data sources. We commence in Section~\ref{sec:gsqd} with a general discussion of quantum sources with graph structure and the design of appropriate loss functions for their characterization. This discussion is then followed in Section~\ref{sec:qnn} with the description of a training algorithm for a quantum neural network ansatz. The results of this algorithm's numerical investigations are then presented in Section~\ref{sec:results}, where unsupervised and semi-supervised scenarios are considered. The main contributions of this paper are: (i) the design of information-theoretic loss functions to capture the graph structure of quantum data sources; (ii) the development of (quantum) training algorithms applicable to QNNs to optimize the loss functions as mentioned earlier; and (iii) proof-of-principle numerical simulations of the developed training algorithms.

	\section{Related Work}
Here we briefly review classical and quantum machine learning approaches for learning graph-structured data. The key challenge in this area is to encode graph structure into continuous low-dimensional representations, or embeddings, in order to exploit classical machine learning techniques. Unsupervised methods \cite{perozzideepwalk2014,qiu2017network,liungram2019,khoslanerd2020} train vertex representations or embeddings while preserving the topological structure of the graph. These representations are then exploited for downstream tasks such as missing link, or vertex label, prediction.
Initial investigations of graph-based semi-supervised learning \cite{zhugaussian2003,belkinmanifold2006}  considered the addition of explicit graph-based regularizations such as Laplacian regularization to the supervised loss term. Recently semi-supervised approaches based on graph-convolution networks (GCNs) \cite{kipfsemi2017,velickovic2018graph,hamilton2017inductive,xuhow2019} have exhibited state-of-the-art performance for node classification and graph-classification tasks. Instead of using an additional graph regularization term in the loss function, these methods encode graph structure directly in the latent representations using neighborhood aggregation techniques. 
For a comprehensive overview and comparison of unsupervised and semi-supervised techniques for graph-structured data, we refer the interested reader to \cite{khoslacomparativestudy2019,wusurvey2020}.

There have already been a variety of investigations of QML for graph-structured classical and quantum data. Firstly, quantum algorithms for classical graph-structured data using a quantum generalization of the random walk were presented in \cite{dernbach}. Semantic knowledge graphs were the subject of \cite{ma}, where a sampling-based quantum algorithm was proposed. Another direction where graph structure has played a crucial role is in quantum generalizations of convolutional neural networks \cite{arunachalamGuestColumnSurvey2017,cong2019quantum}. Here tensor networks with a hierarchical structure have been used to study many-body systems. The approach of incorporating a graph structure into a neural-network ansatz was also explored in \cite{verdon}, leading to generalizations of recurrent neural networks and convolutional neural networks. 
	
	\section{Graph-structured quantum data}\label{sec:gsqd}
	Correlations, both spatial and temporal, are ubiquitous throughout the natural sciences. Capturing the relationships implied by correlations is most naturally achieved in terms of graph structure. This section introduces the notion of \emph{graph-structured quantum data}, which is the central object of study in this paper. 
	
	We commence by introducing some notation. We assume that we have access to a quantum system whose kinematics are characterized by a Hilbert space $\mathcal{H}$. There is no harm in assuming that $\mathcal{H}$ is finite-dimensional and comprised of a collection of $m$ qubits, i.e., $\mathcal{H}\cong (\mathbb{C}^{2})^{\otimes m}$. (The extension to infinite dimensions does not present too many difficulties.) We imagine that we have some source, for example, a quantum device from an (untrusted) commercial purveyor, of quantum states \footnote{Recall that a \emph{quantum state} is a \emph{density operator} $\rho$ on a Hilbert space $\mathcal{H}$, namely a positive semidefinite operator with unit trace: $\rho\ge 0$ and $\text{tr}(\rho) = 1$.} for the quantum system: the source produces an (uncharacterised or untrusted) quantum state $\rho$ on demand. The quantum state produced by the device is assumed distributed according to some probability distribution over a set $\mathcal{S} = \{\rho_{v_1}, \rho_{v_2}, \ldots, \rho_{v_n}\}$ of possible quantum states. Thus we write $\{(p_{v_j}, \rho_{v_j})\}_{j=1}^n$ for the source. So far, this is completely general and characterizes both unstructured and structured quantum data. 
	
	To go further we introduce a graph structure on the quantum data as follows. Suppose that the quantum states $\rho_v$ are associated with the nodes of a graph $G=(V,E)$, i.e., we introduce a map
	\begin{equation}
		\rho:V\rightarrow \mathcal{D}(\mathcal{H})
	\end{equation}
	from the vertices/nodes of the graph $G$ to the set of density operators on $\mathcal{H}$. The \emph{connectivity} structure of the data is captured by the edge set $E$ and quantifies the \emph{information-theoretic closeness}, or \emph{correlations}, between neighboring states. That is, two states $\rho_v$ and $\rho_w$ are \emph{neighboring} with corresponding edge $(v,w)\in E$, if they are close according to an information metric, i.e., $d(\rho_v, \rho_w) \sim \epsilon$. (We discuss the choice and design of the precise information metric below.)
	
	To gain some intuition for this definition, we consider three examples. The first concerns a quantum simulation device which is claimed to simulate some interesting quantum system with Hamiltonian $H \in \mathcal{B}(\mathcal{H})$ for some period of time $t\in  \{0,\epsilon, 2\epsilon, \ldots (n-1)\epsilon\}$. That is, we have quantum states \footnote{The corresponding density operator is $\rho_t\equiv |\psi_t\rangle\langle\psi_t|$} $|\psi_t\rangle \equiv e^{it H}|0\rangle$, where $|0\rangle\in \mathcal{H}$ is some fiducial initial state. Here we associate the path graph $P_n$ on $n$ vertices with this dataset; the vertices label the time associated with $|\psi_t\rangle$. The second example also concerns many-body physics: here, we presume a commercial vendor has produced a quantum device that can supposedly prepare a many-body system into a state with a particle localized at a given position in a lattice (the picture to have in mind here is that of a scanning tunneling microscope). Now the output states $\rho_v$ are labeled by locations on a lattice graph $G$. The third example pertains to irregular graphs with a distribution of vertex degrees and connectivity, namely, a quantum device which emits low-energy eigenstates of \emph{disordered} quantum systems such as \emph{Sachdev-Ye-Kitaev}-type (SYK) models, which have recently received considerable attention in the high-energy physics literature in the context of holography \cite{maldacenaRemarksSachdevYeKitaevModel2016}.
		
	Given graph-structured quantum data $\{(p_v, \rho_v)\}_{v\in V}$, where $\rho_v$ occurs with probability $p_v$, we turn to the goal of learning and modeling the (network of) quantum information processors. (Notice that our graph-based loss functions can be used with many kinds of graph structure. The structure can for example only describe the input states of the network, or the desired outputs. We use the latter case in our numerics in Sec.~\ref{sec:results}.) We assume that the uncharacterized quantum information processor(s) are described by a completely general completely positive (CP) map $\mathcal{F}$ (this CP map provides the complete description of the entire network of processors) \footnote{CP maps are the most general operations allowed in quantum mechanics, see, e.g., \cite{nielsenQuantumComputationQuantum2000} for further discussion.}. The graph structure manifests itself on the \emph{outputs} of the processor(s) $\sigma_v = \mathcal{F}(\rho_v)$. Because two inputs $\rho_v$ and $\rho_w$ which are physically close (i.e., associated with neighbouring vertices) should lead to correlated results when processed by $\mathcal{F}$ we assume that the output states are \emph{information-theoretically} close, written $\sigma_v \sim \sigma_w$. Quantifying and exploiting this information-theoretic closeness is the main goal of this paper.

	Putting aside the precise learning architecture (be it a QAOA or QNN or something completely different) for the moment, we focus first on motivating and defining physically meaningful success metrics. To begin this discussion, we simply assume that our learning architecture is described by a variational class $\mathcal{V}$ of completely positive CP maps  $\mathcal{E}:\mathcal{D}(\mathcal{H}_{\text{in}})\rightarrow \mathcal{D}(\mathcal{H}_{\text{out}})$ which take a quantum state $\rho$ associated with a vertex and process it into some posterior output state $\mathcal{E}(\rho)$. We explain the loss functions we use to train and, after training, test our network in the following subsections.
	
	\subsection{Supervised loss}
	At first we focus on how to subject a subset of the vertices of the graph to supervision. To simplify the description of the loss function in this case we assume that the supervised vertices are required to be pure states (this restriction can be lifted with a little work). In this case the training dataset comprises a list of pairs of pure states:  
	\begin{equation}\label{in_ssv}
		\lbrace \left(\rho_1,\ket{\phi^\text{sv}_1}\bra{\phi^\text{sv}_1}\right),\ldots,\left(\rho_S,\ket{\phi^\text{sv}_S}\bra{\phi^\text{sv}_S}\right),\rho_{S+1},\ldots,\rho_N\rbrace,
	\end{equation}
	where, without loss of generality, we have listed the $S$ supervised (labeled) vertices first followed by the $N-S$ unsupervised (unlabeled) vertices. 
	
	The key operational input required to build a meaningful success metric, or loss function, is a way to measure the information distance between two arbitrary quantum states $\rho$ and $\sigma$. Here the \emph{fidelity}  $F(\rho,\sigma) \equiv \text{tr}(\sqrt{\sqrt{\rho}\sigma \sqrt{\rho}})$ is the natural choice \cite{nielsenQuantumComputationQuantum2000}. The supervised part of our loss function is then
	\begin{equation}\label{eq:Costssv}
		\mathcal{L}_\text{SV} \equiv \frac{1}{S}\sum\limits_{u=1}^S\bra{\phi^\text{sv}_u}\mathcal{E}\left(\rho^\text{in}_u\right)\ket{\phi^\text{sv}_u}
	\end{equation}  
	
	\subsection{Graph-based loss}
	The supervised states are pure, however, the output states of our network are, in general, mixed. Although the fidelity is also defined for mixed states, the excessive computational complexity required to evaluate it metric means that it is often convenient to instead exploit the \emph{Hilbert-Schmidt} distance
	\begin{equation}
		d_{\text{HS}}(\rho,\sigma) \equiv \text{tr}((\rho-\sigma)^2).
	\end{equation} 
	To say that the learning architecture $\mathcal{E}$ has correctly captured the graph structure $G$ of the source and supplied us with a \emph{faithful embedding} we introduce the following loss function
	\begin{equation}
		\mathcal{L}_{G} \equiv \sum_{v,w\in V} [A]_{vw} d_{\text{HS}}(\mathcal{E}(\rho_v),\mathcal{E}(\rho_w)),
	\end{equation}
	where $A$ is the adjacency matrix of the graph $G$ and $[A]_{vw}$ denotes the matrix element of $A$ corresponding to vertices $v$ and $w$. 
	
%
	This loss function is minimized precisely when the processed output states of neighboring vertices in the graph are mapped to informationally close states. 
	
	\subsection{Training loss}
	The full loss function is now specified as the combination of supervised and graph-based loss, with the graph part controlled by a Lagrange multiplier $\gamma$:
	\begin{equation}
		\mathcal{L}_\text{SV+G}=\mathcal{L}_\text{SV} + \gamma \mathcal{L}_{G}.
	\end{equation}   
	The training task is thus to \emph{maximize} $\mathcal{L}_\textit{SV+G}$ with $\gamma \leq 0$. (Recall that two quantum states are closest when the fidelity $F(\rho,\sigma)$ is maximum.) Generically the maximum depends on $\gamma$. In particular, by tuning $\gamma$, one can weight the importance of the graph structure.
	
	It is important here to stress the role played by the graph-based loss $\mathcal{L}_G$: in a semi-supervised learning setting $\mathcal{L}_G$ provides the core mechanism which allows the QNN to \emph{interpolate} between supervised vertices. If $\mathcal{L}_G$ were \emph{not} present, then the QNN would have no mechanism to exploit the graph structure to interpolate the action of $\mathcal{E}$ on unobserved vertices. 

	A crucial feature of our loss function is that it is agnostic of the QNN architecture $\mathcal{E}$: it applies equally to any variational ansatz from QAOA to dissipative QNNs.
	
	\subsection{Testing loss}
	The \emph{testing dataset} is supplied as a complete list of input and output states, containing both the supervised output states as well as the output states which were so far hidden from the QNN:
	\begin{equation}\label{eq:trainingloss}
		\lbrace \left(\rho_1,\ket{\phi^\text{sv}_1}\bra{\phi^\text{sv}_1}\right),\ldots,\left(\rho_N,\ket{\phi^\text{sv}_N}\bra{\phi^\text{sv}_N}\right)\rbrace.
	\end{equation}
	
	After training the network with the loss function Eq.~(\ref{eq:trainingloss}), it is important to check how well the network generalizes, and this means how well it predicts the unsupervised outcomes. We use the following testing loss for this task.
	\begin{equation*}
		\mathcal{L}_\text{USV}=\frac{1}{N-S}\sum_{x=S+1}^{N} \langle\phi^{\text{out}}_x\rvert\rho_x^{\text{out}}\lvert\phi^{\text{out}}_x\rangle.
	\end{equation*}

	\section{A quantum neural network ansatz to learn graph-structured quantum data}\label{sec:qnn}
	We are particularly interested in scenarios where the input and output Hilbert spaces have different dimensions, which captures scenarios from classification through to device characterisation. This is most flexibly modelled via the \emph{dissipative} variational quantum neural network ansatz based on \cite{beerTrainingDeepQuantum2020}. (Note that this QNN ansatz is universal for quantum computation so that it can equally model unitary processes along with general CP maps.) A more detailed description can be found in Appendix~\ref{appendix:review}. 
	
\begin{figure}
		\begin{center}
			\begin{tikzpicture}
				\begin{scope}[xshift=0.9cm,yshift=1.45cm]
					\draw[decorate,decoration={brace,amplitude=3pt}] 
					(-1.25,0) node(t_k_unten){} -- 
					(1.5,0) node(t_k_opt_unten){};   
					\node at (0.1,0.5){$U^1=U_3^1U_2^1U_1^1$};
				\end{scope}
				\foreach \x in {-.5,.5} {
					\draw[white,line width=3pt] (0,\x) -- (2,-1);
					\draw[color2,line width=0.9pt] (0,\x) -- (2,-1);
					\draw[white,line width=3pt] (0,\x) -- (2,0);
					\draw[color1,line width=0.9pt] (0,\x) -- (2,0);
					\draw[white,line width=3pt] (0,\x) -- (2,1);
					\draw[color3,line width=0.9pt] (0,\x) -- (2,1);
				}
				\foreach \x in {-1.5,-0.5, ..., 1.5} {
					\draw[white,line width=3pt] (2,-1) -- (4,\x);
					\draw[line width=0.5pt] (2,-1) -- (4,\x);
					\draw[white,line width=3pt] (2,0) -- (4,\x);
					\draw[line width=0.5pt] (2,0) -- (4,\x);
					\draw[white,line width=3pt] (2,1) -- (4,\x);
					\draw[line width=0.5pt] (2,1) -- (4,\x);
				}
				\foreach \x in {-1.5,-0.5, ..., 1.5} {
					\draw[white,line width=3pt] (4,\x) -- (6,-0.5);
					\draw[line width=0.5pt] (4,\x) -- (6,-0.5);
					\draw[white,line width=3pt] (4,\x) -- (6,0.5);
					\draw[line width=0.5pt] (4,\x) -- (6,0.5);
				}
				\foreach \x in {-1,0,1} {
					\draw[draw,fill=white] (2,\x) circle (7pt);
				}
				\foreach \x in {-1.5,-0.5, ..., 1.5} {
					\draw[fill=white] (4,\x) circle (7pt);
				}
				\draw[fill=white] (0,-0.5) circle (7pt);
				\draw[fill=white] (0,0.5) circle (7pt);
				\draw[fill=white] (6,-0.5) circle (7pt);
				\draw[fill=white] (6,0.5) circle (7pt);
				\node at (0,-2.0){$l=\text{in}$};
				\node at (2,-2.0){$l=1$};
				\node at (3,-2.0){$\cdots$};
				\node at (4,-2.0){$l=L$};
				\node at (6,-2.0){$l=\text{out}$};
				\draw[decorate,decoration={brace,amplitude=3pt,mirror}] 
				(-.75,-2.6) node(t_k_unten){} -- 
				(.75,-2.6) node(t_k_opt_unten){}; 
				\node at (.2,-3){input layer};
				\draw[decorate,decoration={brace,amplitude=3pt,mirror}] 
				(1,-2.6) node(t_k_unten){} -- 
				(5,-2.6) node(t_k_opt_unten){}; 
				\node at (3,-3){hidden layers};
				\draw[decorate,decoration={brace,amplitude=3pt,mirror}] 
				(5.25,-2.6) node(t_k_unten){} -- 
				(6.75,-2.6) node(t_k_opt_unten){}; 
				\node at (5.7,-3){output layer};
			\end{tikzpicture}
			\caption{A quantum feed-forward neural network (QNN) with an \emph{input}, an \emph{output}, and $L$ \emph{hidden} layers. The order of application of the perceptron unitaries is indicated with over/under crossings.}
			\label{fig:genQNN}
		\end{center}
	\end{figure}
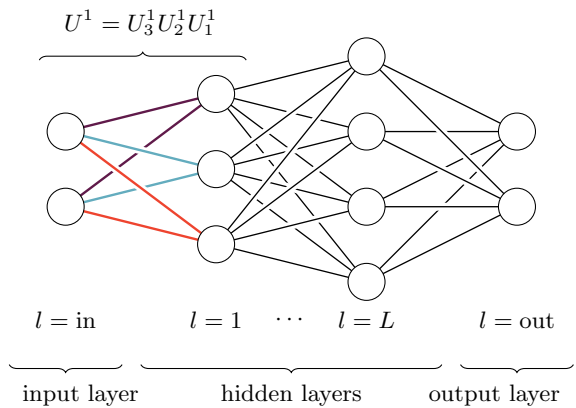

	The QNN ansatz is built from \emph{quantum perceptrons}, which are \emph{general} unitary operators $U$ acting simultaneously on the input and output qubits. The input qubits are assumed in a state $\rho^\mathrm{in}$ and the output qubits in a product state $\lvert 0\dots 0\rangle$. The output of one layer of perceptrons is then 
	\begin{equation}
		\label{eq:Qperceptron}
		\rho^\mathrm{out}=\mathrm{tr}_\mathrm{in}\left(U_\mathrm{in,out}\left(\rho^\mathrm{in}\otimes\lvert 0\dots 0\rangle_\mathrm{out}\langle 0\dots 0\rvert\right)U_\mathrm{in,out}^\dagger\right),
	\end{equation}
	where $U_\mathrm{in,out}$ is the product of all unitaries in that layer. We concentrate, for simplicity, on the case where the quantum perceptrons act on several input qubits and only a \emph{single} output qubit. The general QNN is then described as follows: it consists of an input layer, $L$ hidden layers, and an output layer. See Fig.~\ref{fig:genQNN} for an illustration.
	
	The QNN is a special class of quantum circuit comprised only of quantum perceptrons: the output state of the QNN with $L$ hidden layers is then given by
	\begin{equation}
		\label{eq:rhoout}
		\rho^\mathrm{out}=\mathrm{tr}_\mathrm{in,hidden}\left(U^\mathrm{out}U^L\dots U^1\left(\rho^\mathrm{in}\otimes\lvert 0\dots 0\rangle_\mathrm{hidden,out}\langle 0\dots 0\rvert\right){U^1}^\dagger\dots{U^L}^\dagger {U^\mathrm{out}}^\dagger\right),
	\end{equation}
	where $U^l$ are the layer unitaries, which are comprised of a product of quantum perceptrons acting on the qubits in layer $l-1$ and $l$:
	\begin{equation*}
		U^l=U_{m_l}^lU_{m_l-1}^l\dots U_1^l,
	\end{equation*} 
	where $m_l$ is the number of qubits in layer $l$. 
	
	Since a quantum perceptron is an arbitrary unitary operator, the perceptrons do not, in general, commute. (This is indicated in the figures with over- and under-crossings.) Nevertheless, QNNs still inherit many of the crucial properties of classical NNs. Most particularly, the network output is given by the composition of a sequence of completely positive layer-to-layer \emph{transition maps} $\mathcal{E}^l$:
	\begin{align*}
		\rho_x^\mathrm{out}&=\mathcal{E}_s^{\mathrm{out}}\left(\mathcal{E}^{L}\left(\dots\mathcal{E}^{2}\left(\mathcal{E}^{1}\left(\rho_x^\mathrm{in}\right)\right)\dots\right)\right)
	\end{align*}
	with the channel going from layer $l-1$ to $l$ being
	\begin{align}
		\label{eq:channel}
		\mathcal{E}^{l}\left(X^{l-1}\right)&=\mathrm{tr}_{l-1}\left(U_{m_l}^l\dots U_1^l\left(X^{l-1}\otimes\lvert 0\dots 0\rangle_l\langle 0\dots 0\rvert\right){U_1^l}^\dagger\dots{U_{m_l}^l}^\dagger\right),
	\end{align}
	where $m_l$ is the number of perceptrons in layer $l$.

	With the loss functions and QNN ansatz in hand, we can explain how training proceeds. To optimize the loss function, we exploit gradient descent by allowing the perceptron unitaries to depend on a parameter $s$. We then update the component unitaries of the QNN by the following procedure:
	\begin{equation}
		\label{eq:updateU}
		U_j^l(s+\epsilon)=e^{i\epsilon K_j^l(s)} U_j^l(s).
	\end{equation}
	Here $K_j^l(s)$ are hermitian matrices that are chosen to optimize the loss function.
	The update matrix for a QNN trained with pure states $\lvert\phi^\mathrm{sv}_u\rangle$ as supervised vertices (and without using any known graph structure) is
	\begin{equation}
		K^l_j(s) = \frac{2^{m_{l-1}}i}{S\gamma}\sum\limits_u\mathrm{tr}_\text{rest}\big\{M^l_{j\{u\}}(s)\big\},
	\end{equation}
	where 
	\begin{align*}
		M_{j\{u\}}^l(s)&=\left[U_j^l(s)U_{j-1}^l(s)\dots U_1^1(s)\ \left(\rho_u^\mathrm{in}\otimes\lvert 0\dots 0\rangle_1\langle 0\dots 0\rvert\right) {U_1^1}^\dagger(s)\dots{U_{j-1}^l}^\dagger(s){U_j^l}^\dagger(s),\right.\\
		&\hspace{15pt}\left.{U_{j+1}^l}^\dagger(s)\dots {U_{m_\mathrm{out}}^\mathrm{out}}^\dagger(s)\left(\mathbb{I}_\mathrm{in,hidden}\otimes\lvert\phi^\mathrm{sv}_u\rangle\langle\phi^\mathrm{sv}_u\rvert\right)U_{m_\mathrm{out}}^\mathrm{out}(s)\dots U_{j+1}^l(s)\right].
	\end{align*}
	This is shown in \cite{beerTrainingDeepQuantum2020}.
	
\begin{figure}
		\centering
		\begin{tikzpicture}[node distance=2cm,>=stealth',bend angle=30,auto]
			
			\begin{scope}
				
				\tikzstyle{place}=[circle,draw,fill=color4,minimum size=6mm]
				
				\node [place, fill=color1, pin={[pin distance=7mm,pin edge={dashed,black}]90:$\{\rho^\textit{in}_1,\rho^\textit{sv}_1\}$}] (v1)      {$v_1$};

				\node [place, pin={[pin distance=7mm,pin edge={dashed,black}]270:$\rho^\textit{in}_2$}] (v2) [right of=v1]   {$v_2$}
				edge[color2,line width=0.9pt] node[swap] {$1$} (v1);
				
				\node [above of=v2, yshift=1cm] {\textbf{Graph}};
				
				\node [place, pin={[pin distance=7mm,pin edge={dashed,black}]90:$\rho^\textit{in}_3\in \mathcal{D}(\mathbb{C}^2\otimes \mathbb{C}^2)$}] (v3)  [right of=v2]  {$v_3$}	edge[color2,line width=0.9pt] node[swap] {$1$} (v2);
				
				\node (connector1)  [draw=color1, right of=v2, yshift=1.35cm, xshift=0.11cm, minimum size=5mm]  {};
				
				\node (connector2)  [draw=color1, right of=v2, yshift=1.35cm, xshift=0.97cm, minimum size=5mm]  {};
				
			\end{scope}
			
			\begin{scope}[xshift=8cm, yshift=1cm]
				
				\tikzstyle{place}=[circle,draw,minimum size=6mm]
				
				\node [place] (p1)      {$\mathbb{C}^2$}
				edge [pre, color1, bend right] (connector1.north);
				
				\node [above of=p1, xshift=1cm] {\textbf{QNN}};
				
				\node [place] (p2) [below of=p1]      {$\mathbb{C}^2$}
				edge [pre, color1, bend left] (connector2.south);
				
				\node [place] (p3) [right of=p1, label={[color3]160:$U^1_1$}]   {$\mathbb{C}^2$}
				edge [color3,line width=0.9pt] (p1)
				edge [color3,line width=0.9pt] (p2);
				
				\node [place] (p4) [right of=p2, label={[color3]230:$U^1_2$}]   {$\mathbb{C}^2$};
				
				\draw[white,line width=4pt] (p1) --  (p4) -- (p2);
				
				\draw[color3,line width=0.9pt] (p1) --  (p4) -- (p2);
				
				\node [below of=p1, yshift=1cm, xshift=-0.5cm] {input layer};
				
				\node [below of=p3, yshift=1cm, , xshift=0.5cm] {output layer};
				
			\end{scope}
			
		\end{tikzpicture}
		
		\caption{Illustration of the semi-supervised learning of a graph-structured quantum source (supervised nodes are shaded) via a QNN.}
		\label{fig:pic2}
	\end{figure}
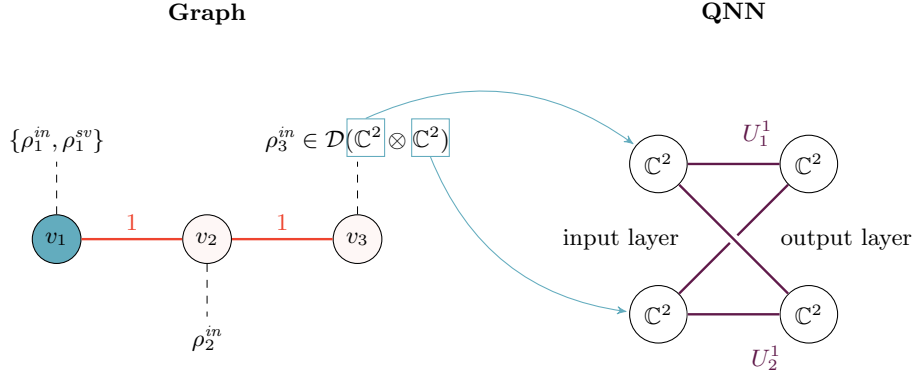

	To explain how the QNN treats graph-structured quantum data and processes, see Fig.~\ref{fig:pic2}: here, we have depicted the graph structure on the left (a path graph on three vertices) and the QNN on the right. Note, particularly, that the topology of the QNN need not have anything to do with the graph structure of the source. Here, the source states are all input states for the QNN and belong to the set of density operators on two qubits. Supervised vertices (in this case one) are shaded, and the corresponding supervised input and output are displayed as a pair $(\rho^\textit{in}_i,\rho^\textit{sv}_i)$.
	
	\begin{thm}
		The update matrix for a QNN trained with a graph structure between output states  $\{\rho^\mathrm{out}_v,\rho^\mathrm{out}_w\}$ encoded in a adjacency matrix $[A]_{vw}$ (and without any supervised states) is
		\begin{equation}
			K^l_j(s)=\frac{2^{m_{l-1}+1}i}{\gamma}\sum\limits_{v\sim w}[A]_{vw}\mathrm{tr}_\text{rest}\big\{M^l_{j\{v,w\}}(s)\big\},
		\end{equation}
		where 
		\begin{align*}
			M_{j\{v,w\}}^l(s)&=\left[U_j^l(s)U_{j-1}^l(s)\dots U_1^1(s)\ \left(\left(\rho_v^\mathrm{in}-\rho_w^\mathrm{in}\right)\otimes\lvert 0\dots 0\rangle_1\langle 0\dots 0\rvert\right) {U_1^1}^\dagger(s)\dots{U_{j-1}^l}^\dagger(s){U_j^l}^\dagger(s),\right.\\
			&\hspace{15pt}\left.{U_{j+1}^l}^\dagger(s)\dots {U_{m_\mathrm{out}}^\mathrm{out}}^\dagger(s)\left(\mathbb{I}_\mathrm{in,hidden}\otimes\left(\rho^\mathrm{out}_v-\rho^\mathrm{out}_w\right)\right)U_{m_\mathrm{out}}^\mathrm{out}(s)\dots U_{j+1}^l(s)\right].
		\end{align*}
	\end{thm}
	See the Appendix~\ref{appendix:derivation} for the proof.
	
	\begin{cor}	
		For a QNN trained with supervised vertices, as well as with graph structure, the update matrix is
		\begin{equation}
			K^l_j(s) = \frac{2^{m_{l-1}}i}{S\gamma}\sum\limits_u\mathrm{tr}_\text{rest}\big\{M^l_{j\{u\}}(s)\big\} + \lambda \frac{2^{m_{l-1}+1}i}{\gamma}\sum\limits_{v\sim w}[A]_{vw}\mathrm{tr}_\text{rest}\big\{M^l_{j\{v,w\}}(s)\big\}.
		\end{equation}
	\end{cor}
	The expression for the update matrices is involved, however, they exhibit a particularly striking structure: one can calculate the updates iteratively, layer by layer, retaining only the reduced state for two layers at a time. This is reminiscent of the update rules arising in the backpropagation algorithm for classical feed-forward neural networks. 

	\section{Results and discussion}\label{sec:results}
	This section describes the results of numerical pilot studies for the semi-supervised learning of graph-structured quantum sources on QNN, with and without the use of graph structure. The aim here is to demonstrate that there are cases where the usage of the graph information leads to better training of the QNN.
	
	These pilot studies were carried out using an exact simulation of the quantum systems on a classical computer. Due to the exponential scaling of the Hilbert space dimension with qubit number, we were limited to small quantum systems. Note, however, that the learning algorithms described here give rise to, with appropriate standard modifications described, e.g., in \cite{beerTrainingDeepQuantum2020} to scalable quantum algorithms suitable for execution on the next generation of quantum computing devices. These quantum algorithms will be described in a future publication.

	\subsection{Example I: connected clusters}
\begin{figure}[h!]
		\begin{center}
				\begin{tikzpicture}[scale=.95]
					\node[place, fill=color4] at (-1,0) (a1) {$v_1$};
					\node[font=\fontsize{8}{0}\selectfont] at ([shift={(180:1.8)}]a1) (a1l) {$\{\ket{\phi_1^\text{in}},\ket{0}\}$};
					\draw[dashed,   black] (a1) -- (a1l.east);
					\node[place, fill=color1]   at (0,1)  (a2)  {$v_2$}; 
					\node[font=\fontsize{8}{0}\selectfont]  at ([shift={(180:3)}]a2) (a2l) {$\{\ket{\phi_2^\text{in}},0.997\ket{0}+0.071\ket{1}\}$};
					\draw[dashed,   black] (a2) -- (a2l.east);
					\node[place, fill=color4]   at (0,-1)  (a3) {$v_3$};
					\node[font=\fontsize{8}{0}\selectfont]  at ([shift={(180:3)}]a3) (a3l) {$\{\ket{\phi_3^\text{in}},0.988\ket{0}+0.152\ket{1}\}$};
					\draw[dashed,   black] (a3) -- (a3l.east);
					\node[place, fill=color1] at (1,0) (a4) {$v_4$};
					\node[font=\fontsize{8}{0}\selectfont]  at ([shift={(-15:2.8)}]a4) (a4l) {$\{\ket{\phi_4^\text{in}},0.97\ket{0}+0.243\ket{1}\}$};
					\draw[dashed,   black] (a4) -- (a4l.north west);
					\node[place, fill=color4]  at (4,0) (b1) {$v_8$};
					\node[font=\fontsize{8}{0}\selectfont]  at ([shift={(90:1.2)}]b1) (b1l) {$\{\ket{\phi_8^\text{in}},0.659\ket{0}+0.753\ket{1}\}$};
					\draw[dashed,   black] (b1) -- (b1l.south);
					\node[place, fill=color1] at (8,0) (c1) {$v_5$};
					\node[font=\fontsize{8}{0}\selectfont]  at ([shift={(0:3)}]c1) (c1l) {$\{\ket{\phi_5^\text{in}},0.152\ket{0}+0.988\ket{1}\}$};
					\draw[dashed,   black] (c1) -- (c1l.west);
					\node[place, fill=color4] at (9,1) (c2) {$v_6$};
					\node[font=\fontsize{8}{0}\selectfont]  at ([shift={(0:3)}]c2) (c2l) {$\{\ket{\phi_6^\text{in}},0.071\ket{0}+0.997\ket{1}\}$};
					\draw[dashed,   black] (c2) -- (c2l.west);
					\node[place, fill=color4] at (9,-1) (c3) {$v_7$};
					\node[font=\fontsize{8}{0}\selectfont]  at ([shift={(0:1.8)}]c3) (c3l) {$\{\ket{\phi_7^\text{in}},\ket{1}\}$};
					\draw[dashed,   black] (c3) -- (c3l.west);
					\draw[color2,line width=0.9pt] (a1) -- (a2);		
					\draw[color2,line width=0.9pt] (a1) -- (a3);
					\draw[color2,line width=0.9pt] (a1) -- (a4);
					\draw[color2,line width=0.9pt] (a2) -- (a3);
					\draw[color2,line width=0.9pt] (a2) -- (a4);
					\draw[color2,line width=0.9pt] (a3) -- (a4);
					\draw[color2,line width=0.9pt] (a4) --  (b1) -- (c1);
					\draw[color2,line width=0.9pt] (c1) -- (c2);		
					\draw[color2,line width=0.9pt] (c1) -- (c3);
					\draw[color2,line width=0.9pt] (c2) -- (c3);
				\end{tikzpicture}
			\caption{\textbf{Connected clusters: Data I.} The output states of this training dataset comprise a graph with two clusters. The states are chosen in a way such that $v_1\hdots v_4$ form one cluster, $v_8$ connects the two clusters, and the remaining three vertices form the second cluster. The coefficients are only recorded to three decimal places. Notice that only $S$ of these states are used for training. (In the figure $S=3$ example supervised vertices are shaded.) The input states are generally taken to be unstructured. In our case the states $\ket{\phi_i^\text{in}}$ are random $3$-qubit states built via a normal (Gaussian) distribution. }
			\label{connectedClusters}
		\end{center}
	\end{figure}
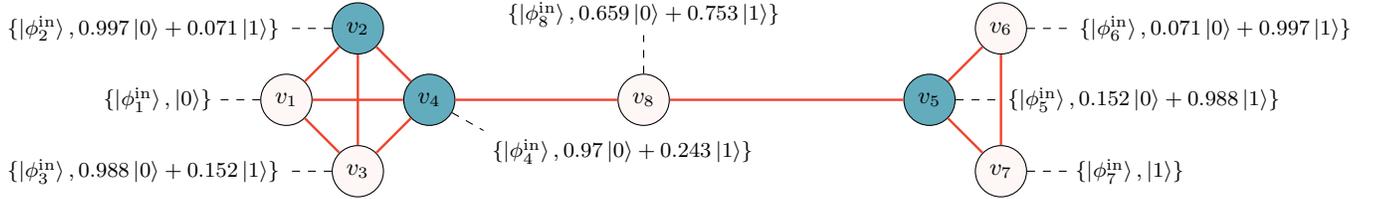

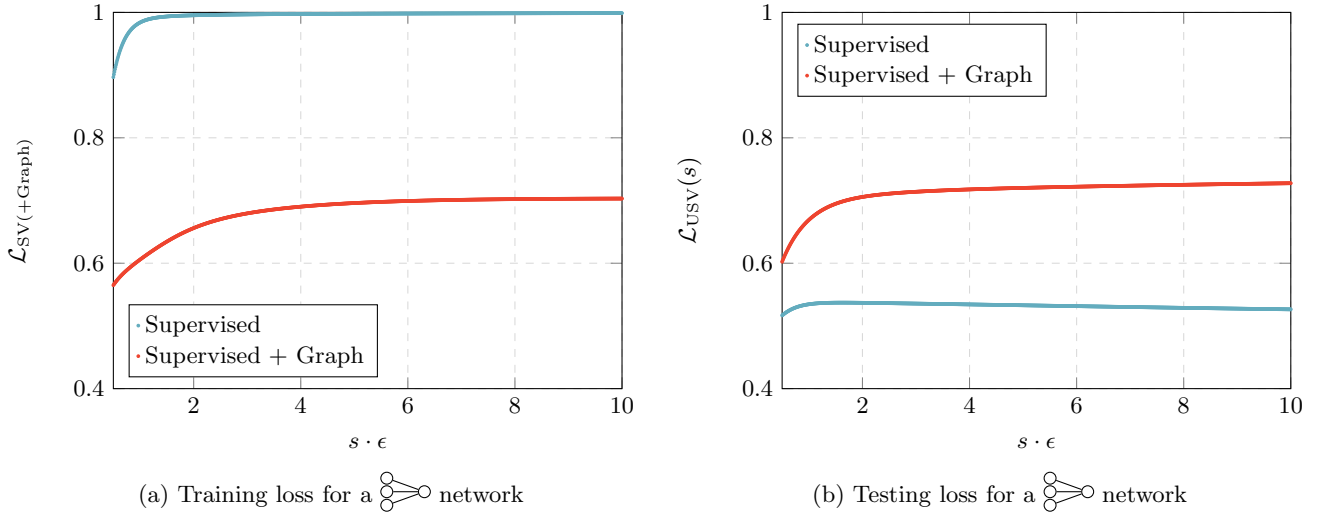
\begin{figure}[H]
		\centering
		\begin{subfigure}{0.49\textwidth}
			\begin{tikzpicture}
				\begin{axis}[
					xmin=0.5,   xmax=10,
					ymin=0.4,   ymax=1,
					width=.95\linewidth, 
					height=.75\linewidth,
					grid=major,
					grid style={dashed,gray!30},
					xlabel= $s \cdot \epsilon$, 
					ylabel=$\mathcal{L}_\text{SV(+Graph)}$,legend pos=south west,legend cell align={left}]
					\addplot[color=color1, only marks, mark size=0.5pt,mark phase=0] table [x=step times epsilon, y=SsvTraining, col sep=comma] {connectedClustersRandomShuffled_8pairs3sv_3-1network_adjT0i65and1_g-0i5_delta0_lda1_ep0i01_plot.csv};
					\addlegendentry{Supervised} 
					\addplot[color=color2, only marks, mark size=0.5pt,mark phase=0] table [x=step times epsilon, y=SsvGraphTraining, col sep=comma] {connectedClustersRandomShuffled_8pairs3sv_3-1network_adjT0i65and1_g-0i5_delta0_lda1_ep0i01_plot.csv};
					\addlegendentry{Supervised + Graph} 
				\end{axis}
			\end{tikzpicture}
			\subcaption{Training loss for a \begin{tikzpicture}[yscale=0.6,xscale=0.5, baseline=-4]
					\node(1) [circle,draw,inner sep=0pt,minimum size=4.5pt] at (0,-.3) {};
					\node(2) [circle,draw,inner sep=0pt,minimum size=4.5pt] at (0,0) {};
					\node(3) [circle,draw,inner sep=0pt,minimum size=4.5pt] at (0,.3) {};
					\node(4) [circle,draw,inner sep=0pt,minimum size=4.5pt] at (1,0) {};
					\draw (1)--(4);
					\draw (2)--(4);
					\draw (3)--(4);
				\end{tikzpicture} network}
		\end{subfigure}
		\begin{subfigure}{0.49\textwidth}
			\begin{tikzpicture}
				\begin{axis}[
					xmin=0.5,   xmax=10,
					ymin=0.4,   ymax=1,
					width=.95\linewidth, 
					height=.75\linewidth,
					grid=major,
					grid style={dashed,gray!30},
					xlabel= $s \cdot \epsilon$, 
					ylabel=$\mathcal{L}_\text{USV}(s)$,legend pos=north west,legend cell align={left}]
					\addplot[color=color1, only marks, mark size=0.5pt,mark phase=0] table [x=step times epsilon, y=SsvTestingUsv, col sep=comma] {connectedClustersRandomShuffled_8pairs3sv_3-1network_adjT0i65and1_g-0i5_delta0_lda1_ep0i01_plot.csv};
					\addlegendentry{Supervised} 
					\addplot[color=color2, only marks, mark size=0.5pt,mark phase=0] table [x=step times epsilon, y=SsvGraphTestingUsv, col sep=comma] {connectedClustersRandomShuffled_8pairs3sv_3-1network_adjT0i65and1_g-0i5_delta0_lda1_ep0i01_plot.csv};
					\addlegendentry{Supervised + Graph} 
				\end{axis}
			\end{tikzpicture}
			\subcaption{Testing loss for a \begin{tikzpicture}[yscale=0.6,xscale=0.5, baseline=-4]
					\node(1) [circle,draw,inner sep=0pt,minimum size=4.5pt] at (0,-.3) {};
					\node(2) [circle,draw,inner sep=0pt,minimum size=4.5pt] at (0,0) {};
					\node(3) [circle,draw,inner sep=0pt,minimum size=4.5pt] at (0,.3) {};
					\node(4) [circle,draw,inner sep=0pt,minimum size=4.5pt] at (1,0) {};
					\draw (1)--(4);
					\draw (2)--(4);
					\draw (3)--(4);
				\end{tikzpicture} network}
			\label{connected clusters training b}
		\end{subfigure}
		\caption{\textbf{Connected clusters: training the QNN.} We trained the network with three supervised training pairs from Fig.~\ref{connectedClusters} and the there-presented graph structure. The training (a) and the testing loss (b) during $1000$ rounds of training ($\epsilon=0.01$) with $\gamma=0$ semi-supervised (green) and  $\gamma=-0.5$ semi-supervised plus graph information (blue) is plotted.}
		\label{connected clusters training}
	\end{figure}

	For the first numerical study, we construct a graph of $N=8$ pairs of quantum states in the form of two connected clusters. The resulting structure is depicted in Fig.~\ref{connectedClusters}. 
	The evolution of the loss functions during the training is depicted in Fig.~\ref{connected clusters training}. One can easily observe that the network performs better during the testing procedure: see Fig.~\ref{connected clusters training b}, where the graph structure was exploited during training.

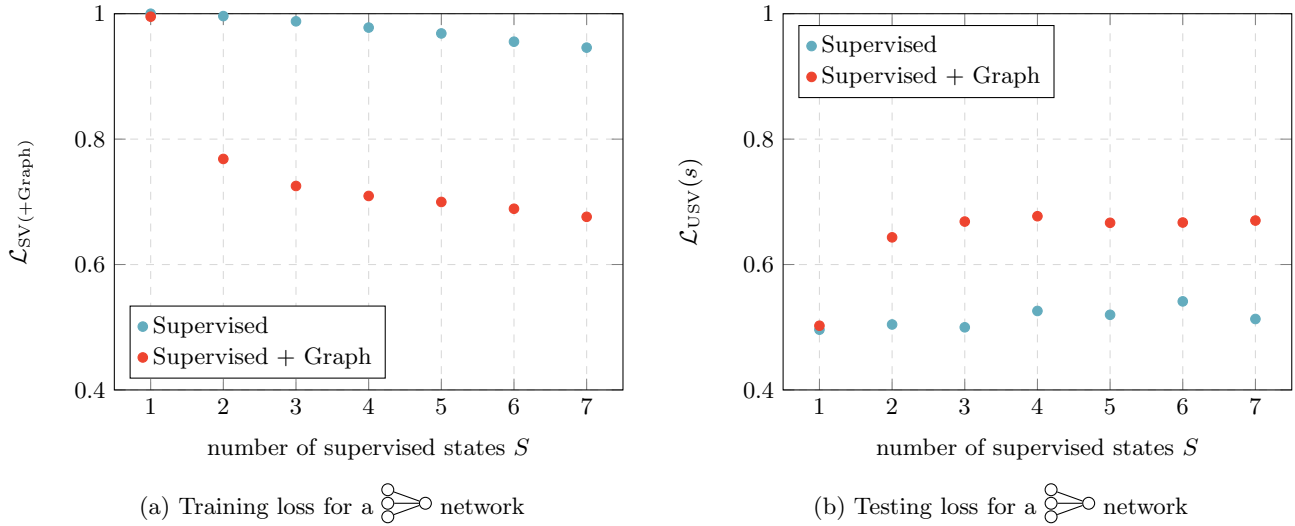
\begin{figure}[H]
		\centering
		\begin{subfigure}{0.49\textwidth}
			\begin{tikzpicture}
				\begin{axis}[
					xmin=0.5,   xmax=7.5,
					ymin=0.4,   ymax=1,
					width=.95\linewidth, 
					height=.75\linewidth,
					grid=major,
					grid style={dashed,gray!30},
					xlabel=number of supervised states $S$, 
					ylabel=$\mathcal{L}_\text{SV(+Graph)}$,legend pos=south west,legend cell align={left}]
					\addplot[color=color1, only marks, mark size=1.8pt,mark phase=0] table [x=numberSupervisedPairsList, y=SsvTrainingMeanList, col sep=comma] {connectedClustersRandomShuffled_8pairs_3-1network_adjT0i65and1_g-0i5_delta0_lda1_ep0i01_rounds1000_shots30_plotmean.csv};
					\addlegendentry{Supervised} 
					\addplot[color=color2, only marks, mark size=1.8pt,mark phase=0] table [x=numberSupervisedPairsList, y=SsvGraphTrainingMeanList, col sep=comma] {connectedClustersRandomShuffled_8pairs_3-1network_adjT0i65and1_g-0i5_delta0_lda1_ep0i01_rounds1000_shots30_plotmean.csv};
					\addlegendentry{Supervised + Graph} 
				\end{axis}
			\end{tikzpicture}
			\subcaption{Training loss for a \begin{tikzpicture}[yscale=0.6,xscale=0.5, baseline=-4]
					\node(1) [circle,draw,inner sep=0pt,minimum size=4.5pt] at (0,-.3) {};
					\node(2) [circle,draw,inner sep=0pt,minimum size=4.5pt] at (0,0) {};
					\node(3) [circle,draw,inner sep=0pt,minimum size=4.5pt] at (0,.3) {};
					\node(4) [circle,draw,inner sep=0pt,minimum size=4.5pt] at (1,0) {};
					\draw (1)--(4);
					\draw (2)--(4);
					\draw (3)--(4);
			\end{tikzpicture} network}
		\end{subfigure}
		\begin{subfigure}{0.49\textwidth}
			\begin{tikzpicture}
				\begin{axis}[
					xmin=0.5,   xmax=7.5,
					ymin=0.4,   ymax=1,
					width=.95\linewidth, 
					height=.75\linewidth,
					grid=major,
					grid style={dashed,gray!30},
					xlabel=number of supervised states $S$, 
					ylabel=$\mathcal{L}_\text{USV}(s)$,legend pos=north west,legend cell align={left}]
					\addplot[color=color1, only marks, mark size=1.8pt,mark phase=0] table [x=numberSupervisedPairsList, y=SsvTestingUsvMeanList, col sep=comma] {connectedClustersRandomShuffled_8pairs_3-1network_adjT0i65and1_g-0i5_delta0_lda1_ep0i01_rounds1000_shots30_plotmean.csv};
					\addlegendentry{Supervised} 
					\addplot[color=color2, only marks, mark size=1.8pt,mark phase=0] table [x=numberSupervisedPairsList, y=SsvGraphTestingUsvMeanList, col sep=comma] {connectedClustersRandomShuffled_8pairs_3-1network_adjT0i65and1_g-0i5_delta0_lda1_ep0i01_rounds1000_shots30_plotmean.csv};
					\addlegendentry{Supervised + Graph} 
				\end{axis}
			\end{tikzpicture}
			\subcaption{Testing loss for a \begin{tikzpicture}[yscale=0.6,xscale=0.5, baseline=-4]
					\node(1) [circle,draw,inner sep=0pt,minimum size=4.5pt] at (0,-.3) {};
					\node(2) [circle,draw,inner sep=0pt,minimum size=4.5pt] at (0,0) {};
					\node(3) [circle,draw,inner sep=0pt,minimum size=4.5pt] at (0,.3) {};
					\node(4) [circle,draw,inner sep=0pt,minimum size=4.5pt] at (1,0) {};
					\draw (1)--(4);
					\draw (2)--(4);
					\draw (3)--(4);
				\end{tikzpicture} network}
		\end{subfigure}
		\caption{\textbf{Connected clusters: training for different numbers of supervised pairs $S$.} We trained the network with Data I. The training (a) and the testing loss (b) after $1000$ rounds of training ($\epsilon=0.01$) with $\gamma=0$ semi-supervised (green) and $\gamma=-0.5$ semi-supervised plus graph information (blue) averaged over $10$ shots is plotted for different $S$ (supervised pairs).}
		\label{connected clusters plot}
	\end{figure}

	In the first experiment in Fig.~\ref{connected clusters training} three of the eight vertices were supervised. We have studied how the number of supervised vertices affects the training process. Therefore we randomly chose $S<N$ of the $N=8$ training pairs to be supervised before every training shot and trained the \begin{tikzpicture}[yscale=0.6,xscale=0.5, baseline=-4]
		\node(1) [circle,draw,inner sep=0pt,minimum size=4.5pt] at (0,-.3) {};
		\node(2) [circle,draw,inner sep=0pt,minimum size=4.5pt] at (0,0) {};
		\node(3) [circle,draw,inner sep=0pt,minimum size=4.5pt] at (0,.3) {};
		\node(4) [circle,draw,inner sep=0pt,minimum size=4.5pt] at (1,0) {};
		\draw (1)--(4);
		\draw (2)--(4);
		\draw (3)--(4);
	\end{tikzpicture} network for $1000$ training rounds. After $30$ shots, we built the mean of the loss value after training. These training and the testing loss means are displayed against $S\in\{0,N\}$ in Fig.~\ref{connected clusters plot}.

	As may be readily observed from the figures, the QNN is able to interpolate the action of the learned operation on unobserved vertices. One may also observe the trade-off between the test loss in the cases with and without graph structure.

	\subsection{Example II: Line}
\begin{figure}[h!]
		\begin{center}
				
				\begin{tikzpicture}[scale=1.8]
					\node[place, fill=color4] at (0,0) (a1) {$v_1$};
					\node[font=\fontsize{8}{0}\selectfont] at ([shift={(90:0.7)}]a1) (a1l) {$\{\ket{\phi_1^\text{in}},\ket{0}\}$};
					\draw[dashed,   black] (a1) -- (a1l);
					\node[place, fill=color1]   at (1,0) (a2)  {$v_2$}; 
					\node[font=\fontsize{8}{0}\selectfont] at ([shift={(270:0.7)}]a2) (a2l) {$\{\ket{\phi_2^\text{in}},0.99\ket{0}+0.21\ket{1}\}$};
					\draw[dashed,   black] (a2) -- (a2l);
					\node[place, fill=color1]   at (2,0) (a3) {$v_3$};
					\node[font=\fontsize{8}{0}\selectfont] at ([shift={(90:0.7)}]a3) (a3l) {$\{\ket{\phi_3^\text{in}},0.96\ket{0}+0.28\ket{1}\}$};					
					\draw[dashed,   black] (a3) -- (a3l);
					\node[place, fill=color4] at (3,0) (a4) {$v_4$};
					\node[font=\fontsize{8}{0}\selectfont] at ([shift={(270:0.7)}]a4) (a4l) {$\{\ket{\phi_4^\text{in}},0.89\ket{0}+0.45\ket{1}\}$};					
					\draw[dashed,   black] (a4) -- (a4l);
					\node[place, fill=color4]  at (4,0) (a5) {$v_5$};
					\node[font=\fontsize{8}{0}\selectfont] at ([shift={(90:0.7)}]a5) (a5l) {$\{\ket{\phi_5^\text{in}},0.78\ket{0}+0.62\ket{1}\}$};					
					\draw[dashed,   black] (a5) -- (a5l);
					\node[place, fill=color4] at (5,0) (a6) {$v_6$};
					\node[font=\fontsize{8}{0}\selectfont] at ([shift={(270:0.7)}]a6) (a6l) {$\{\ket{\phi_6^\text{in}},0.62\ket{0}+0.78\ket{1}\}$};					
					\draw[dashed,   black] (a6) -- (a6l);
					\node[place, fill=color4] at (6,0) (a7) {$v_7$};
					\node[font=\fontsize{8}{0}\selectfont] at ([shift={(90:0.7)}]a7) (a7l) {$\{\ket{\phi_7^\text{in}},0.45\ket{0}+0.89\ket{1}\}$};					
					\draw[dashed,   black] (a7) -- (a7l);
					\node[place, fill=color1] at (7,0) (a8) {$v_8$};
					\node[font=\fontsize{8}{0}\selectfont] at ([shift={(270:0.7)}]a8) (a8l) {$\{\ket{\phi_8^\text{in}},0.27\ket{0}+0.96\ket{1}\}$};					
					\draw[dashed,   black] (a8) -- (a8l);
					\node[place, fill=color4] at (8,0) (a9) {$v_9$};
					\node[font=\fontsize{8}{0}\selectfont] at ([shift={(90:0.7)}]a9) (a9l) {$\{\ket{\phi_9^\text{in}},0.12\ket{0}+0.99\ket{1}\}$};					
					\draw[dashed,   black] (a9) -- (a9l);
					\node[place, fill=color4] at (9,0) (a10) {$v_{10}$};
					\node[font=\fontsize{8}{0}\selectfont] at ([shift={(270:0.7)}]a10) (a10l) {$\{\ket{\phi_{10}^\text{in}},\ket{1}\}$};					
					\draw[dashed,   black] (a10) -- (a10l);
					\draw[color2,line width=0.9pt] (a1) -- (a2) -- (a3) -- (a4) -- (a5) -- (a6) -- (a7) -- (a8) -- (a9) -- (a10);
					
				\end{tikzpicture}
			\caption{\textbf{Line: Data II.} the output states of this training dataset form a line graph. The states were chosen to be -- according to the fidelity -- evenly spaced along a line between the states $\ket{0}$ and $\ket{1}$ associated with the endpoints. Again, notice that only $S$ of these states are used for training. (Here $S=3$ example vertices are shaded.) The input states are again unstructured: $\ket{\phi_i^\text{in}}$ are random $3$-qubit states built from a normal (Gaussian) distribution. }
				\label{line}
			\end{center}
		\end{figure}
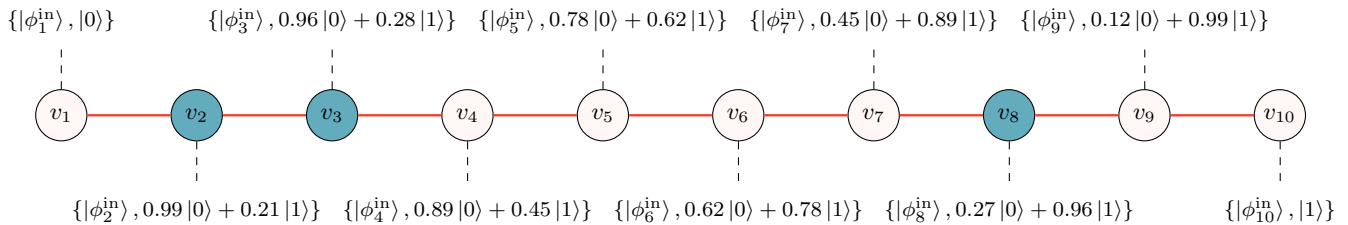

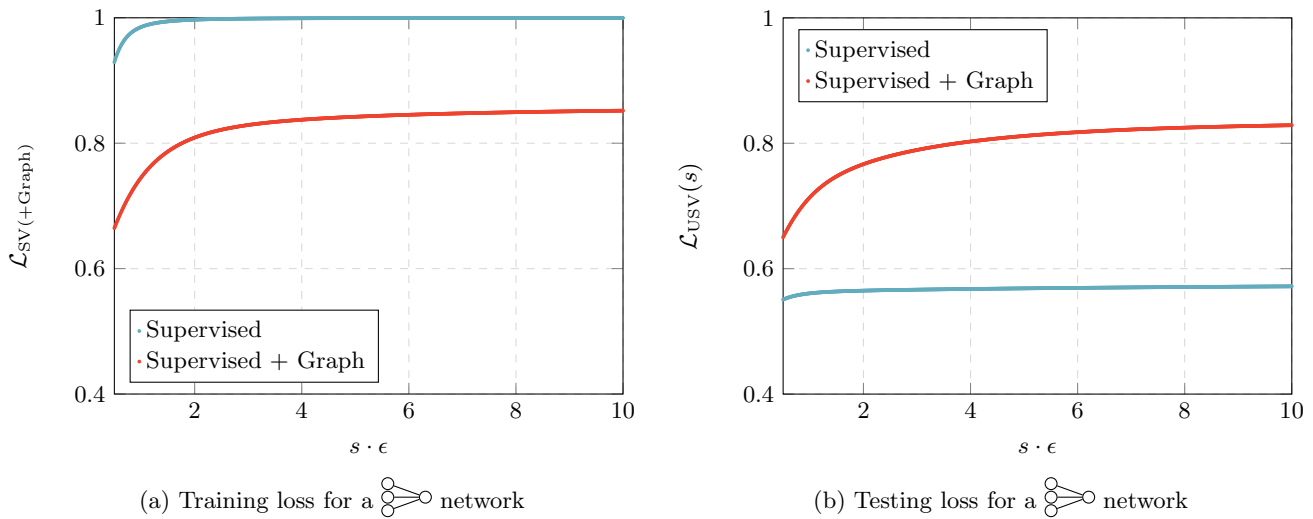
\begin{figure}[H]
			\centering
			\begin{subfigure}{0.49\textwidth}
				\begin{tikzpicture}
					\begin{axis}[
						xmin=0.5,   xmax=10,
						ymin=0.4,   ymax=1,
						width=.95\linewidth, 
						height=.75\linewidth,
						grid=major,
						grid style={dashed,gray!30},
						xlabel= $s \cdot \epsilon$, 
						ylabel=$\mathcal{L}_\text{SV(+Graph)}$,legend pos=south west,legend cell align={left}]
						\addplot[color=color1, only marks, mark size=0.5pt,mark phase=0] table [x=step times epsilon, y=SsvTraining, col sep=comma] {lineOutputRandomShuffled_10pairs3sv_3-1network_adjT0i93and1_g-0i5_delta0_lda1_ep0i01_plot.csv};
						\addlegendentry{Supervised} 
						\addplot[color=color2, only marks, mark size=0.5pt,mark phase=0] table [x=step times epsilon, y=SsvGraphTraining, col sep=comma] {lineOutputRandomShuffled_10pairs3sv_3-1network_adjT0i93and1_g-0i5_delta0_lda1_ep0i01_plot.csv};
						\addlegendentry{Supervised + Graph} 
					\end{axis}
				\end{tikzpicture}
				\subcaption{Training loss for a \begin{tikzpicture}[yscale=0.6,xscale=0.5, baseline=-4]
						\node(1) [circle,draw,inner sep=0pt,minimum size=4.5pt] at (0,-.3) {};
						\node(2) [circle,draw,inner sep=0pt,minimum size=4.5pt] at (0,0) {};
						\node(3) [circle,draw,inner sep=0pt,minimum size=4.5pt] at (0,.3) {};
						\node(4) [circle,draw,inner sep=0pt,minimum size=4.5pt] at (1,0) {};
						\draw (1)--(4);
						\draw (2)--(4);
						\draw (3)--(4);
					\end{tikzpicture} network}
			\end{subfigure}
			\begin{subfigure}{0.49\textwidth}
				\begin{tikzpicture}
					\begin{axis}[
						xmin=0.5,   xmax=10,
						ymin=0.4,   ymax=1,
						width=.95\linewidth, 
						height=.75\linewidth,
						grid=major,
						grid style={dashed,gray!30},
						xlabel= $s \cdot \epsilon$, 
						ylabel=$\mathcal{L}_\text{USV}(s)$,legend pos=north west,legend cell align={left}]
						\addplot[color=color1, only marks, mark size=0.5pt,mark phase=0] table [x=step times epsilon, y=SsvTestingUsv, col sep=comma] {lineOutputRandomShuffled_10pairs3sv_3-1network_adjT0i93and1_g-0i5_delta0_lda1_ep0i01_plot.csv};
						\addlegendentry{Supervised} 
						\addplot[color=color2, only marks, mark size=0.5pt,mark phase=0] table [x=step times epsilon, y=SsvGraphTestingUsv, col sep=comma] {lineOutputRandomShuffled_10pairs3sv_3-1network_adjT0i93and1_g-0i5_delta0_lda1_ep0i01_plot.csv};
						\addlegendentry{Supervised + Graph} 
					\end{axis}
				\end{tikzpicture}
				\subcaption{Testing loss for a \begin{tikzpicture}[yscale=0.6,xscale=0.5, baseline=-4]
						\node(1) [circle,draw,inner sep=0pt,minimum size=4.5pt] at (0,-.3) {};
						\node(2) [circle,draw,inner sep=0pt,minimum size=4.5pt] at (0,0) {};
						\node(3) [circle,draw,inner sep=0pt,minimum size=4.5pt] at (0,.3) {};
						\node(4) [circle,draw,inner sep=0pt,minimum size=4.5pt] at (1,0) {};
						\draw (1)--(4);
						\draw (2)--(4);
						\draw (3)--(4);
					\end{tikzpicture} network}
			\end{subfigure}
			\caption{\textbf{Line: training the QNN.} We trained the network with three supervised training pairs of the states from Fig.~\ref{line} and the line graph structure. The training (a) and the testing loss (b) during $1000$ rounds of training ($\epsilon=0.01$) with $\gamma=0$ semi-supervised (green) and $\gamma=-1$ semi-supervised plus graph information (blue) is plotted.}
			\label{line training}
		\end{figure}

\begin{figure}[H]
			\centering
			\begin{subfigure}{0.49\textwidth}
				\begin{tikzpicture}
					\begin{axis}[
						xmin=0.5,   xmax=9.5,
						ymin=0.4,   ymax=1,
						width=.95\linewidth, 
						height=.75\linewidth,
						grid=major,
						grid style={dashed,gray!30},
						xlabel=number of supervised states $S$, 
						ylabel=$\mathcal{L}_\text{SV(+Graph)}$,legend pos=south west,legend cell align={left}]
						\addplot[color=color1, only marks, mark size=1.8pt,mark phase=0] table [x=numberSupervisedPairsList, y=SsvTrainingMeanList, col sep=comma] {lineOutputRandomShuffled_10pairs_3-1network_adjT0i93and1_g-0i5_delta0_lda1_ep0i01_rounds1000_shots30_plotmean.csv};
						\addlegendentry{Supervised} 
						\addplot[color=color2, only marks, mark size=1.8pt,mark phase=0] table [x=numberSupervisedPairsList, y=SsvGraphTrainingMeanList, col sep=comma] {lineOutputRandomShuffled_10pairs_3-1network_adjT0i93and1_g-0i5_delta0_lda1_ep0i01_rounds1000_shots30_plotmean.csv};
						\addlegendentry{Supervised + Graph} 
					\end{axis}
				\end{tikzpicture}
				\subcaption{Training loss for a \begin{tikzpicture}[yscale=0.6,xscale=0.5, baseline=-4]
						\node(1) [circle,draw,inner sep=0pt,minimum size=4.5pt] at (0,-.3) {};
						\node(2) [circle,draw,inner sep=0pt,minimum size=4.5pt] at (0,0) {};
						\node(3) [circle,draw,inner sep=0pt,minimum size=4.5pt] at (0,.3) {};
						\node(4) [circle,draw,inner sep=0pt,minimum size=4.5pt] at (1,0) {};
						\draw (1)--(4);
						\draw (2)--(4);
						\draw (3)--(4);
					\end{tikzpicture} network}
			\end{subfigure}
			\begin{subfigure}{0.49\textwidth}
				\begin{tikzpicture}
					\begin{axis}[
						xmin=0.5,   xmax=9.5,
						ymin=0.4,   ymax=1,
						width=.95\linewidth, 
						height=.75\linewidth,
						grid=major,
						grid style={dashed,gray!30},
						xlabel=number of supervised states $S$, 
						ylabel=$\mathcal{L}_\text{USV}(s)$,legend pos=north west,legend cell align={left}]
						\addplot[color=color1, only marks, mark size=1.8pt,mark phase=0] table [x=numberSupervisedPairsList, y=SsvTestingUsvMeanList, col sep=comma] {lineOutputRandomShuffled_10pairs_3-1network_adjT0i93and1_g-0i5_delta0_lda1_ep0i01_rounds1000_shots30_plotmean.csv};
						\addlegendentry{Supervised} 
						\addplot[color=color2, only marks, mark size=1.8pt,mark phase=0] table [x=numberSupervisedPairsList, y=SsvGraphTestingUsvMeanList, col sep=comma] {lineOutputRandomShuffled_10pairs_3-1network_adjT0i93and1_g-0i5_delta0_lda1_ep0i01_rounds1000_shots30_plotmean.csv};
						\addlegendentry{Supervised + Graph} 
					\end{axis}
				\end{tikzpicture}
				\subcaption{Testing loss for a \begin{tikzpicture}[yscale=0.6,xscale=0.5, baseline=-4]
						\node(1) [circle,draw,inner sep=0pt,minimum size=4.5pt] at (0,-.3) {};
						\node(2) [circle,draw,inner sep=0pt,minimum size=4.5pt] at (0,0) {};
						\node(3) [circle,draw,inner sep=0pt,minimum size=4.5pt] at (0,.3) {};
						\node(4) [circle,draw,inner sep=0pt,minimum size=4.5pt] at (1,0) {};
						\draw (1)--(4);
						\draw (2)--(4);
						\draw (3)--(4);
					\end{tikzpicture} network}
			\end{subfigure}
			\caption{\textbf{Line: Training for different numbers of supervised pairs $S$.} We train the network with Data II. The training (a) and the testing loss (b) after $1000$ rounds of training $\epsilon=0.01$ with $\gamma=0$ semi-supervised (green) $\gamma=-1$ semi-supervised plus graph information (blue) averaged from $30$ shots is plotted for different number of $S$ (supervised pairs).}
			\label{line plot}
		\end{figure}
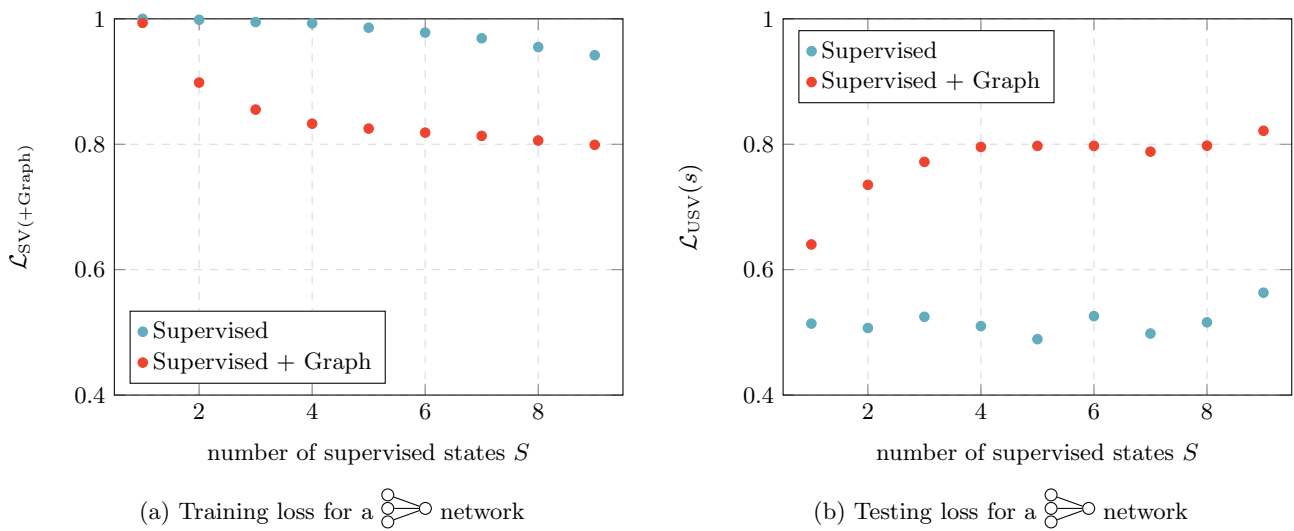

	For the second example we choose $N=10$ pairs of quantum states chosen so that the correlation structure is encoded in a line graph, see Fig.~\ref{line}.  
	As in Example I, we plotted the training and testing loss for one training in Fig.~\ref{line training}. We also varied the number of supervised states $S<10$. The results are displayed in Fig.~\ref{line plot}.

		In this example, one observes that one can achieve a testing loss of over $0.6$ with only $5$ of $10$ supervised vertices when the graph structure is exploited. These numerical results demonstrate that graph structure information provides powerful side information for training.

		\section*{Conclusions}
		In this paper, we have considered the learning of graph-structured quantum sources using dissipative QNNs. We have explained how to exploit the graph structure by designing information-theoretic loss functions. The optimization of the loss functions via QNNs was described, leading to analytic formulas for the update rules. Finally, proof-of-principle numerical simulations of the developed training algorithms were carried out, demonstrating the remarkable ability of trained QNNs to interpolate between supervised vertices and infer unobserved vertex labellings. 		
		
		\section*{Acknowledgments}
		Helpful correspondence and discussions with Dmytro Bondarenko, Terry Farrelly, Polina Feldmann, Alexander Hahn, Gabriel Müller, Jan Hendrik Pfau, Robert Salzmann, Daniel Scheiermann, Viktoria Schmiesing, Marvin Schwiering, Christian Struckmann and Ramona Wolf are gratefully acknowledged. This work was supported, in part, by the Quantum Valley Lower Saxony (QVLS), the DFG through SFB 1227 (DQ-mat), the RTG 1991, and funded by the Deutsche Forschungsgemeinschaft (DFG, German Research Foundation) under Germany’s Excellence Strategy EXC-2123 QuantumFrontiers 390837967, and LeibnizKILabor (01DD20003) funded by the BMBF.

		\section*{Data availability:} All results were obtained using Python. The code is available at 
		
		\url{https://github.com/qigitphannover/DeepQuantumNeuralNetworks}.
	
		\bibliographystyle{ieeetr}
		\bibliography{paper}
		
		\newpage
			\section{Appendix}
	\label{appendix:review}
	
	\subsection{Classical simulation of training the QNN}
	\label{sec:TrainingQNN}
	All numerical results we present in this paper were obtained using Python. The details are described in Fig.~\ref{algorithm}.
	
	\begin{figure}[h!]
		\begin{center}
			\adjustbox{minipage=0.97\textwidth,cfbox=color3}{
				\begin{minipage}{0.97 \textwidth}
					\begin{itemize}
						\item[I.] Initialization:
						\begin{itemize}
							\item[I.1] Set $s=0$.
							\item[I.2] Choose the networks unitaries $U_j^l(0)$ randomly.
						\end{itemize}
						\item[II.] Feed forward: The following steps are equivalent to applying the layer-to-layer channels $\mathcal{E}_s^{l}$ defined in Equation~\ref{eq:channel} successively to the input state. For each $\left(\lvert\phi^\mathrm{in}_x\rangle,\lvert\phi^\mathrm{out}_x\rangle\right)$ in the set of training data and for every layer $l$, do the following:
						\begin{itemize}
							\item[II.1] Tensor the state of the layer to the output state of layer $l-1$, where $\rho_x^\mathrm{in}=\lvert\phi^\mathrm{in}_x\rangle\langle\phi^\mathrm{in}_x\rvert$:
							\begin{equation*}
								\rho_x^{l-1}(s)\otimes\lvert 0\dots 0\rangle_l\langle 0\dots 0\rvert
							\end{equation*}
							\item[II.2] Apply the unitaries in layer $l$:
							\begin{equation*}
								U_{m(l)}^l(s) U_{m(l)-1}^l(s)\dots U_1^l(s)\left(\rho_x^{l-1}(s)\otimes\lvert 0\dots 0\rangle_l\langle 0\dots 0\rvert\right) {U_1^l}^\dagger(s)\dots{U_{m(l)-1}^l}^\dagger(s) {U_{m(l)}^l}^\dagger(s)
							\end{equation*}
							\item[II.3] Trace out layer $l-1$:
							\begin{equation*}
								\rho_x^l(s)=\mathrm{tr}_{l-1}\left(U_{m(l)}^l(s) U_{m(l)-1}^l(s)\dots U_1^l(s)\left(\rho_x^{l-1}(s)\otimes\lvert 0\dots 0\rangle_l\langle 0\dots 0\rvert\right) {U_1^l}^\dagger(s)\dots{U_{m(l)-1}^l}^\dagger(s) {U_{m(l)}^l}^\dagger(s)\right)
							\end{equation*}
							\item[II.4] Store $\rho_x^l(s)$. This information is needed to compute the parameter matrices.
						\end{itemize}
						\item[III.] Update parameters:
						\begin{itemize}
							\item[III.1] Compute the training loss: 
							\begin{equation*}
								C(s)=\frac{1}{N}\sum_{x=1}^N\langle\phi^\mathrm{out}_x\rvert\rho_x^\mathrm{out}(s)\lvert\phi^\mathrm{out}_x\rangle
							\end{equation*}
							\item[III.2] Calculate each parameter matrix $K_j^l(s)$.
							\item[III.3] Update each perceptron unitary via
							\begin{equation*}
								U_j^l(s+\epsilon)=e^{i\epsilon K_j^l(s)}U_j^l(s).
							\end{equation*}
							\item[III.4] Update $s=s+\epsilon$.
						\end{itemize}
						\item[IV.] Repeat steps II. and III. until the training loss has reached its maximum.
					\end{itemize}
				\end{minipage}
			}
		\end{center}
		\caption{\textbf{Algorithm:} Classical simulation of the QNN.}
		\label{algorithm}
	\end{figure}

	\subsection{Derivation of update rules}
	\label{appendix:derivation}
	As explained above, our QNN update rule for $j$th qubit in $l$th layer is determined by a Hermitian matrix $K^l_j$ . A good choice of such an update matrix is one that cause a decrease (or increase, depending on the task) in the evaluated loss at each training step. The change in loss with respect to step parameter is given as
	\begin{equation*}
		\frac{d\mathcal{L}(s)}{ds}=\lim\limits_{\epsilon \rightarrow 0}\frac{\mathcal{L}(s+\epsilon)-\mathcal{L}(s)}{\epsilon}.
	\end{equation*}
	In what follows we  denote $\left(\rho^\mathrm{in}_i\otimes \ket{0,\ldots}_\mathrm{in,hidden}\bra{0,\ldots}\right)$ (see Step II.2 in Fig.~\ref{algorithm}) by $\tilde{\rho}_i$.
	For brevity, we leave out the argument $s$ for $U^l_m(s)$ and $K^l_m(s)$ matrices. We first expand $\mathcal{E}_{s+\epsilon}(\rho^\mathrm{in}_i)$ in first order $\epsilon$:
	\begin{eqnarray}
		\mathcal{E}_{s+\epsilon}(\rho^\mathrm{in}_i)&=&\tr_\mathrm{in,hidden}\left(e^{i\epsilon K^\mathrm{out}_{m_\mathrm{out}}}U^\mathrm{out}_{m_\mathrm{out}}\ldots e^{i\epsilon K^1_1}U^1_1\tilde{\rho}_i{U^1_1}^\dagger e^{-i\epsilon K^1_1}\ldots {U^\mathrm{out}_{m_\mathrm{out}}}^\dagger e^{-i\epsilon K^\mathrm{out}_{m_\mathrm{out}}}\right)\nonumber\\
		&=&\mathcal{E}_s(\rho^\mathrm{in}_i) + i\epsilon \tr_\mathrm{in,hidden}\left( K^\mathrm{out}_{m_\mathrm{out}}U^\mathrm{out}_{m_\mathrm{out}}\ldots U^1_1\tilde{\rho}_i{U^1_1}^\dagger\ldots {U^\mathrm{out}_{m_\mathrm{out}}}^\dagger\right.\nonumber\\
		&&- U^\mathrm{out}_{m_\mathrm{out}}\ldots U^1_1 \tilde{\rho}_i {U^1_1}^\dagger \ldots {U^\mathrm{out}_{m_\mathrm{out}}}^\dagger K^\mathrm{out}_{m_\mathrm{out}} +\ldots+ U^\mathrm{out}_{m_\mathrm{out}}\ldots K^1_1 U^1_1\tilde{\rho}_i{U^1_1}^\dagger\ldots {U^\mathrm{out}_{m_\mathrm{out}}}^\dagger\nonumber\\
		&&\left. - U^\mathrm{out}_{m_\mathrm{out}}\ldots U^1_1 \tilde{\rho}_i {U^1_1}^\dagger K^1_1 \ldots {U^\textit{out}_{m_\textit{out}}}^\dagger \right) +\mathcal{O}(\epsilon^2)\nonumber\\
		&=&\rho^\mathrm{out}_i(s) + i\epsilon \tr_\mathrm{in, hidden}\left( \left[K^\mathrm{out}_{m_\mathrm{out}}, U^\mathrm{out}_{m_\mathrm{out}}\ldots U^1_1\tilde{\rho}_i{U^1_1}^\dagger\ldots {U^\mathrm{out}_{m_\mathrm{out}}}^\dagger\right] + \ldots\right.\nonumber\\
		&&\left.+  U^\mathrm{out}_{m_\mathrm{out}}\ldots U^1_2 \left[K^1_1, U^1_1\tilde{\rho}_i{U^1_1}^\dagger\right]{U^1_2}^\dagger\ldots {U^\mathrm{out}_{m_\mathrm{out}}}^\dagger\right) +\mathcal{O}(\epsilon^2)\nonumber\\
		&=& \rho^\mathrm{out}_i(s) + i\epsilon X_s(\rho^\mathrm{in}_i)+\mathcal{O}(\epsilon^2).
	\end{eqnarray}
	\subsubsection{Derivation of supervised update}
	In this subsection, we summarize the derivation of the supervised loss. See \cite{beerTrainingDeepQuantum2020} for more details. 
	
	Taking the derivative of the supervised loss with respect to the step parameter we obtain
	\begin{eqnarray*}
		\frac{d\mathcal{L}_\mathrm{sv}(s)}{ds}&=&\frac{i}{S}\sum\limits_{u=1}^S \tr\left( \ket{\phi^\mathrm{sv}_u}\bra{\phi^\mathrm{sv}_u} X_s( \rho^\mathrm{in}_u)\right)\\
		&=&\frac{i}{S}\sum\limits_{u=1}^S\tr\left( \left(\mathbb{I}_\mathrm{in,hidden}\otimes \ket{\phi^\mathrm{sv}_u}\bra{\phi^\mathrm{sv}_u}\right)\Big(\left[K^\mathrm{out}_{m_\mathrm{out}}, U^\mathrm{out}_{m_\mathrm{out}}\ldots U^1_1\tilde{\rho}_u{U^1_1}^\dagger\ldots {U^\textit{out}_{m_\mathrm{out}}}^\dagger\right]+\ldots\right.\\
		&&\left. + U^\mathrm{out}_{m_\mathrm{out}}\ldots U^1_2 \left[K^1_1, U^1_1\tilde{\rho}_u{U^1_1}^\dagger\right]{U^1_2}^\dagger\ldots {U^\mathrm{out}_{m_\mathrm{out}}}^\dagger\Big)\right)\\
		&=&\frac{i}{S}\sum\limits_{u=1}^S\tr\left( \left[U^\mathrm{out}_{m_\mathrm{out}}\ldots U^1_1\tilde{\rho}_u{U^1_1}^\dagger\ldots {U^\mathrm{out}_{m_\mathrm{out}}}^\dagger,\left(\mathbb{I}_\mathrm{in,hidden}\otimes  \ket{\phi^\mathrm{sv}_u}\bra{\phi^\mathrm{sv}_u}\right)\right]K^\mathrm{out}_{m_\mathrm{out}}+\ldots\right.\\
		&&\left.+ \left[ U^1_1\tilde{\rho}_u{U^1_1}^\dagger,{U^1_2}^\dagger\ldots {U^\mathrm{out}_{m_\mathrm{out}}}^\dagger\left(\mathbb{I}_\mathrm{in,hidden}\otimes \ket{\phi^\mathrm{sv}_u}\bra{\phi^\mathrm{sv}_u} \right)U^\mathrm{out}_{m_\mathrm{out}}\ldots U^1_2 \right] K^1_1\right)\\
		&=&\frac{i}{S}\sum\limits_{u=1}^S\tr\left( M^\mathrm{out}_{m_\mathrm{out} \lbrace u\rbrace}(s)K^\mathrm{out}_{m_\mathrm{out}}(s) +\ldots + M^1_{1 \lbrace u\rbrace}(s)K^1_1(s)\right),
	\end{eqnarray*}
	where
	\begin{align*}
		M_{j\{u\}}^l(s)&=\left[U_j^l(s)U_{j-1}^l(s)\dots U_1^1(s)\ \left(\rho_u^\mathrm{in}\otimes\lvert 0\dots 0\rangle_1\langle 0\dots 0\rvert\right) {U_1^1}^\dagger(s)\dots{U_{j-1}^l}^\dagger(s){U_j^l}^\dagger(s),\right.\\
		&\hspace{15pt}\left.{U_{j+1}^l}^\dagger(s)\dots {U_{m_\mathrm{out}}^\mathrm{out}}^\dagger(s)\left(\mathbb{I}_\mathrm{in,hidden}\otimes\lvert\phi^\mathrm{sv}_u\rangle\langle\phi^\mathrm{sv}_u\rvert\right)U_{m_\mathrm{out}}^\mathrm{out}(s)\dots U_{j+1}^l(s)\right].
	\end{align*}
	Applying the Lagrange method results in
	\begin{eqnarray}
		K^l_{j}(s)&=&\frac{2^{m_{l-1}}i}{S\gamma}\sum\limits_{u=1}^S\tr_\text{rest}\left\{M^l_{j\{u\}}(s)\right\}\in H\left(\mathcal{H}_{l-1}\otimes\mathbb{C}^2\right).
	\end{eqnarray}
	\subsubsection{Derivation of graph-based update}
	Here we prove the following theorem, stated above.  
	\begin{thm}
		The update matrix for a QNN trained with a graph structure between output states  $\{\rho^\mathrm{out}_v,\rho^\mathrm{out}_w\}$ encode via adjacency matrix $[A]_{vw}$ (and without any supervised states) is
		\begin{equation}
			K^l_j(s)=\frac{2^{m_{l-1}+1}i}{\gamma}\sum\limits_{v\sim w}[A]_{vw}\mathrm{tr}_\text{rest}\big\{M^l_{j\{v,w\}}(s)\big\},
		\end{equation}
		where 
		\begin{align*}
			M_{j\{v,w\}}^l(s)&=\left[U_j^l(s)U_{j-1}^l(s)\dots U_1^1(s)\ \left(\left(\rho_v^\mathrm{in}-\rho_w^\mathrm{in}\right)\otimes\lvert 0\dots 0\rangle_1\langle 0\dots 0\rvert\right) {U_1^1}^\dagger(s)\dots{U_{j-1}^l}^\dagger(s){U_j^l}^\dagger(s),\right.\\
			&\hspace{15pt}\left.{U_{j+1}^l}^\dagger(s)\dots {U_{m_\mathrm{out}}^\mathrm{out}}^\dagger(s)\left(\mathbb{I}_\mathrm{in,hidden}\otimes\left(\rho^\mathrm{out}_v-\rho^\mathrm{out}_w\right)\right)U_{m_\mathrm{out}}^\mathrm{out}(s)\dots U_{j+1}^l(s)\right].
		\end{align*}
	\end{thm}

	The graph part of the loss function involves 
	\begin{equation}
		\mathcal{L}_{G} \equiv \sum_{v,w\in V} [A]_{vw} d_{\text{HS}}(\mathcal{E}(\rho_v),\mathcal{E}(\rho_w)),
	\end{equation}
	as cost function. Taking the derivative of $\mathcal{L}_{G}$ with respect to the step parameter we obtain
	\begin{eqnarray}\label{derusv}
		\frac{d\mathcal{L}_G(s)}{ds}&=&\lim\limits_{\epsilon \rightarrow 0}\frac{\mathcal{L}_G(s+\epsilon)-\mathcal{L}_G(s)}{\epsilon}\nonumber\\
		&=&2i\sum\limits_{v,w\in V}[A]_{vw}\tr\left(\left( \rho^\mathrm{out}_v(s)- \rho^\mathrm{out}_w(s)\right)\cdot\left(X_s( \rho^\mathrm{in}_v)-X_s(\rho^\mathrm{in}_w)\right)\right)\nonumber\\
		&=&2i\sum\limits_{v,w\in V}[A]_{vw}\tr\left(\left(\mathbb{I}_\mathrm{in, hidden}\otimes \left( \rho^\mathrm{out}_v- \rho^\mathrm{out}_w\right)\right)\Big(\left[K^\mathrm{out}_{m_\mathrm{out}}, U^\mathrm{out}_{m_\mathrm{out}}\ldots U^1_1\left(\tilde{\rho}_v-\tilde{\rho}_w\right){U^1_1}^\dagger\ldots {U^\mathrm{out}_{m_\mathrm{out}}}^\dagger\right]+\ldots\right. \nonumber\\
		&&\left.+\; U^\mathrm{out}_{m_\mathrm{out}}\ldots U^1_2 \left[K^1_1, U^1_1\left(\tilde{\rho}_i-\tilde{\rho}_j\right){U^1_1}^\dagger\right]{U^1_2}^\dagger\ldots {U^\mathrm{out}_{m_\mathrm{out}}}^\dagger\Big)\right)\nonumber\\
		&=&2i\sum\limits_{v,w\in V}[A]_{vw}\tr\left(\left[U^\mathrm{out}_{m_\mathrm{out}}\ldots U^1_1\left(\tilde{\rho}_v-\tilde{\rho}_w\right){U^1_1}^\dagger\ldots {U^\mathrm{out}_{m_\mathrm{out}}}^\dagger,\left(\mathbb{I}_\mathrm{in, hidden}\otimes \left( \rho^\mathrm{out}_v- \rho^\mathrm{out}_w\right)\right)\right]K^\mathrm{out}_{m_\mathrm{out}}+\ldots\right.\nonumber\\
		&&\left.+ \left[ U^1_1\left(\tilde{\rho}_v-\tilde{\rho}_w\right){U^1_1}^\dagger,{U^1_2}^\dagger\ldots {U^\mathrm{out}_{m_\mathrm{out}}}^\dagger\left(\mathbb{I}_\mathrm{in, hidden}\otimes \left( \rho^\mathrm{out}_v- \rho^\mathrm{out}_w\right)\right)U^\mathrm{out}_{m_\mathrm{out}}\ldots U^1_2 \right] K^1_1\right)\nonumber\\
		&=&2i\sum\limits_{v,w\in V}[A]_{vw}\tr\left( M^\mathrm{out}_{m_\mathrm{out} \lbrace i,j\rbrace}(s)K^\mathrm{out}_{m_\mathrm{out}}(s) +\ldots M^1_{1 \lbrace i,j\rbrace}(s)K^1_1(s)\right)
	\end{eqnarray}
	with
	\begin{align*}
		M_{j\{v,w\}}^l(s)&=\left[U_j^l(s)U_{j-1}^l(s)\dots U_1^1(s)\ \left(\left(\rho_v^\mathrm{in}-\rho_w^\mathrm{in}\right)\otimes\lvert 0\dots 0\rangle_1\langle 0\dots 0\rvert\right) {U_1^1}^\dagger(s)\dots{U_{j-1}^l}^\dagger(s){U_j^l}^\dagger(s),\right.\\
		&\hspace{15pt}\left.{U_{j+1}^l}^\dagger(s)\dots {U_{m_\mathrm{out}}^\mathrm{out}}^\dagger(s)\left(\mathbb{I}_\mathrm{in,hidden}\otimes\left(\rho^\mathrm{out}_v(s)-\rho^\mathrm{out}_w(s)\right)\right)U_{m_\mathrm{out}}^\mathrm{out}(s)\dots U_{j+1}^l(s)\right],
	\end{align*}
	where we used the expansion of $\mathcal{E}_{s+\epsilon}(\rho^\mathrm{in}_i)$.
	Expanding $K^l_j(s)$ via Pauli matrices
	\begin{equation}\label{Kexpand}
		K^l_j(s)=\sum\limits_{\alpha_1,\ldots,\alpha_{m_{l-1}},\beta}K^l_{j,\alpha_1,\ldots,\alpha_{m_{l-1}},\beta}(s)\left( \sigma^{\alpha_1}\otimes\ldots\otimes\sigma^{\alpha_{m_{l-1}}}\otimes\sigma^\beta\right)\in \text{H}\left(\mathcal{H}_{l-1}\otimes\mathbb{C}^2\right),
	\end{equation}
	with coefficients $K^l_{j,\alpha_1,\ldots,\alpha_{m_{l-1}},\beta}(s)\in \mathbb{S}$, then $d_s\mathcal{L}_G(s)$ is linear in these pre-factors. Actually this matrix has to be a Hermitian matrix on the full QNN Hilbert space but without loss of generality, we can reduce our calculations to $\mathcal{H}_{l-1}\otimes\mathbb{C}^2$.
	
	Our way to minimize the cost function is to choose $K^l_j$ in such a way that with every step the cost function decreases, which requires minimizing Equation \eqref{derusv} with respect to \eqref{Kexpand}. The problem is that it reaches its minimum at infinity, to first order in $\epsilon$. To obtain a finite solution we use the Lagrange method with the following condition
	\begin{equation*}
		\sum\limits_{\alpha_1,\ldots,\beta}K_{\alpha_1\ldots,\beta}(s)^2=0.
	\end{equation*}
	So we have to solve
	\begin{eqnarray*}\label{lagrangemin}
		\min\limits_{K^l_{j,\alpha_1,\ldots,\beta}}\left(\frac{d\mathcal{L}_G(s)}{ds}-\gamma\sum\limits_{\alpha_1,\ldots,\beta}K^l_{j,\alpha_1\ldots,\beta}(s)^2\right)
	\end{eqnarray*}
	with the Lagrange multiplier $\gamma$, which has to be chosen in such a way that $d_s\mathcal{L}_G(s)$ is negative. Then we obtain
	\begin{eqnarray*}
		\min\limits_{K^l_{j,\alpha_1,\ldots,\beta}}\Big(2i\sum\limits_{v,w\in V }[A]_{vw}\sum\limits_{\left(l,m\right)}\sum\limits_{\alpha_1,\ldots,\beta}K^l_{j,\alpha_1\ldots,\beta}\tr\left(M^l_{j\{v,w\}}\left( \sigma^{\alpha_1}\otimes\ldots\otimes\sigma^\beta\right)\right)
		-\gamma\sum\limits_{\alpha_1,\ldots,\beta} {K^l_{j,\alpha_1\ldots,\beta}}^2\Big).
	\end{eqnarray*}
	Taking the derivative with respect to $K^l_{j,\alpha_1,\ldots,\beta}$ leads to
	\begin{eqnarray}\label{factors}
		2\gamma K^l_{j,\alpha_1\ldots,\beta}&=&2i\sum\limits_{v,w \in V}[A]_{vw}\tr\left\{M^l_{j\{v,w\}}\left( \sigma^{\alpha_1}\otimes\ldots\otimes\sigma^\beta\right)\right\}.
	\end{eqnarray}
	Then we split the trace, thus we get
	\begin{eqnarray*}
		K^l_{j,\alpha_1\ldots,\beta}&=&\frac{i}{\gamma}\sum\limits_{v,w \in V}[A]_{vw}\tr_{\{\alpha_1,\ldots,\alpha_{m_{l-1}},\beta\}}\left\{\tr_\text{rest}\left\{M^l_{j\{v,w\}}\right\}\left( \sigma^{\alpha_1}\otimes\ldots\otimes\sigma^\beta\right)\right\}.
	\end{eqnarray*}
	Expanding the complex matrix $\tr_\text{rest}\left\{M^l_{j\{v,w\}}\right\}$ with Pauli matrices leads to
	\begin{eqnarray*}
		K^l_{j,\alpha_1\ldots,\beta}&=&\frac{i}{\lambda}\sum\limits_{v,w \in V}[A]_{vw}\tr_{\{\alpha_1,\ldots,\alpha_{m_{l-1}},\beta\}}\left\{\tr_\text{rest}\left\{M^l_{j\{v,w\}}\right\}_{\{\mu_1,\ldots,\nu\}}\left( \sigma^{\mu_1}\otimes\ldots\otimes\sigma^\nu\right)\left( \sigma^{\alpha_1}\otimes\ldots\otimes\sigma^\beta\right)\right\}\\
		&=&\frac{i}{\gamma}\sum\limits_{v,w\in V}[A]_{vw}\tr_\text{rest}\left\{M^l_{j\{v,w\}}\right\}_{\{\mu_1,\ldots,\nu\}}\tr_{\{\alpha_1,\ldots,\alpha_{m_{l-1}},\beta\}}\left\{\left( \sigma^{\mu_1}\otimes\ldots\otimes\sigma^\nu\right)\left( \sigma^{\alpha_1}\otimes\ldots\otimes\sigma^\beta\right)\right\}\\
		&=&\frac{2^{m_{l-1}+1}i}{\gamma}\sum\limits_{v,w\in V}[A]_{vw}\tr_\text{rest}\left\{M^l_{j\{v,w\}}(s)\right\}_{\{\alpha_1,\ldots,\beta\}}.
	\end{eqnarray*}
	This yields the full matrix
	\begin{eqnarray}\label{Kusv}
		K^l_j(s)&=&\frac{2^{m_{l-1}+1}i}{\gamma}\sum\limits_{v,w\in V}[A]_{vw}\sum\limits_{\alpha_1,\ldots,\beta}\tr_\text{rest}\left\{M^l_{j\{v,w\}}(s)\right\}_{\{\alpha_1,\ldots,\beta\}}\left( \sigma^{\alpha_1}\otimes\ldots\otimes\sigma^\beta\right)\nonumber\\
		&=&\frac{2^{m_{l-1}+1}i}{\gamma}\sum\limits_{v,w\in V}[A]_{vw}\tr_\text{rest}\left\{M^l_{j\{v,w\}}(s)\right\}\in H\left(\mathcal{H}_{l-1}\otimes\mathbb{C}^2\right).
	\end{eqnarray}
	As we can see from Equation \eqref{factors}, $d_s\mathcal{L}_G(s) < 0$ for $\gamma < 0$.
	
	\subsubsection{Derivation of full update}
	Since all calculations we have made so far were linear in the cost function, it suffices to compute the update matrix corresponding to supervised loss separately and then combine it with the update matrix for the unsupervised part to obtain the final update matrix as follows
	\begin{equation*}
		K^l_m(s) = \text{supervised update} + \lambda \cdot \text{ unsupervised update matrix}.
	\end{equation*}
	The full semi supervised update matrix is
	\begin{equation}
		K^l_j(s) = \frac{2^{m_{l-1}}i}{S\gamma}\sum\limits_{u=1}^S\tr_\text{rest}\left\{M^l_{j\{u\}}(s)\right\} + \lambda \frac{2^{m_{l-1}+1}i}{\gamma}\sum\limits_{v,w\in V}[A]_{vw}\tr_\text{rest}\left\{M^l_{j\{v,w\}}(s)\right\},
	\end{equation} 
	where $\gamma >0$, as $\lambda<0$ and the cost function should be maximized.
\end{document}